\documentclass{cernyrep}
\usepackage[bookmarks, colorlinks=true, linktoc=page, pdftex, linkcolor=black, 
citecolor=black, urlcolor=black,breaklinks]{hyperref}
\renewcommand\appendixname{Appendix}

\usepackage{fancyhdr}
\fancyhfoffset{4 mm}
\fancypagestyle{ARTTITLE}{%
\fancyhf{} 
\lhead{\small{Proceedings of the 2018 CERN--Accelerator--School course on\\
\it{Numerical Methods for Analysis, Design and Modelling of Particle Accelerators}, Thessaloniki, (Greece)}} 
\lfoot{Available online at \url{https://cas.web.cern.ch/previous-schools}}
\rfoot{\thepage\hspace*{3mm}}
 
}
\begin{document}
\title{Direct Vlasov solvers}
\author{N. Mounet\thanks{nicolas.mounet@cern.ch}}
\institute{CERN, Geneva, Switzerland}


\maketitle
\thispagestyle{ARTTITLE}
\begin{abstract}
In these proceedings we will describe the theory and practical steps required 
to build Vlasov solvers such as those commonly used to compute coherent 
instabilities in synchrotrons. Thanks to a Hamiltonian formalism, we will 
derive a compact and general form of the linearized Vlasov equation, written 
using Poisson brackets. This in turn will be the basis of a procedure 
to build Vlasov solvers, applied to the specific example of transverse 
instabilities arising from beam coupling impedance.
\end{abstract}

\section{Introduction}
\label{sec:intro}

In particle accelerators and storage rings, beam quality is sometimes 
affected by so-called `collective effects', which can lead to e.g. 
emittance growth or intensity loss. These effects are characterized by various 
kinds of interactions between beam particles, building up to a collective 
behaviour of the ensemble of particles which fundamentally differs from that of 
a set of independent particles.

Several such collective effects can lead to coherent instabilities in which the 
beam average position exhibits self-enhanced, growing oscillations. These can 
ultimately lead to a significant decrease of the beam brightness, or even a 
complete loss of the beam in the worst case, and are a source of limitation to 
the operation of particle accelerators. For instance, the transverse 
instabilities observed in the Large Hadron Collider (LHC) during run I and II at 
top energy, led to the use of very high current in the octupolar magnets (close 
to the maximum available) to provide enough Landau 
damping~\cite{ref:EMetral_summary_LHC_instab_review_runI,
ref:EMetral_summary_LHC_instab_review_runII}. Other examples of limitations due 
to beam instabilities in hadron synchrotrons were reviewed in 
Refs.~\cite{ref:IEEE,ref:Salvant_Benevento}.

Coherent instabilities can be caused by various kinds of collective effects
(or a combination of several of them):

\begin{itemize}
\item beam-coupling impedance, i.e. self-generated electromagnetic fields 
obtained through the interaction between the beam and its surroundings (vacuum 
pipe, kickers, collimators, cavities, etc.),
\item an electron cloud around the beam, due to secondary 
emission of electrons at the surface surrounding the beam,
\item interactions with ions trapped in the beam potential,
\item beam-beam effects, i.e. interaction with a counter-rotating beam.
\end{itemize}

There are essentially two common ways to model such interactions:

\begin{itemize}
\item \textit{macroparticle simulations} (or \textit{indirect Vlasov solvers}): 
the beam is looked as a collection of macroparticles which are tracked down the 
full ring. Both the machine optics and the particle interactions are included in
the dynamics of each macroparticle, in an attempt to be as realistic as possible.
The goal is to observe directly the time evolution of the beam transverse motion and 
hence spot possible instabilities. The number of macroparticles is still typically
much smaller than the number of actual particles,
\item \textit{direct Vlasov solvers} (hereafter simply named
\textit{Vlasov solvers}): the phase space distribution is modelled as a 
continuum rather than a collection of discrete particles, and Vlasov equation is 
solved.
\end{itemize}

Historically, the latter approach was the first adopted to try to understand 
the instabilities observed in particle 
accelerators~\cite{ref:Laslett,ref:Sacherer1972}. More recently, 
macroparticles simulations have been used extensively and have shown their 
capabilities to model the beam behaviour in complex situations. Tracking 
macroparticles through such simulations is indeed efficient, often extensible 
at will to almost any kind of particle interactions, and potentially very close 
to reality. We refer the reader to Ref.~\cite{ref:Kli} for a detailed review of 
macroparticles simulations and their numerical implementation.
 
Despite its success in many aspects, this approach may still suffer 
from two possibly important limitations. First, it is a pure time domain technique, and 
simulations need to be stopped after tracking a certain number of turns. Therefore, slow 
instabilities can be missed, and these can be critical in some machines where the beam 
remains stored for a very long time (hours, in the case of the LHC). Second, such 
simulations do not necessarily provide a synthetic understanding of the kind of 
instability, neither of the main parameters nor mitigation means that could prevent it, 
unless large and computationally intensive parameter scans are performed.

Vlasov solvers, on the contrary, are typically limited to simpler
situations, but can provide a good understanding of instability modes and their 
mitigation parameters. Moreover, they are typically very fast. Finally, they are in principle
able to spot an instability irrespectively of its rapidity to develop.

For more detailed reviews on the two approaches and comparisons between 
them, the reader is referred to 
Refs.~\cite{ref:Migliorati,ref:KLi_MSchenk,ref:Mounet_Benevento}. Here we will focus 
on direct Vlasov solvers only, both from the theoretical and the practical 
point of view.

The three main sections of these proceedings are rather independent from each other.
In Section~\ref{sec:Vlasov} we will introduce the 
concept of phase space distribution density and provide Vlasov equation. In 
Section~\ref{sec:linear} we obtain the linearized form of this equation in a compact way, 
\Eref{eq:linear_Vlasov}, as well as higher order extensions, thanks to Poisson brackets 
and a Hamiltonian formalism. The core of these proceedings will then be given in 
Section~\ref{sec:method}, where a method to build a Vlasov solver will 
be described through its application to the specific case of transverse 
instabilities resulting from beam-coupling impedance, for a bunched beam. 
Finally, we will conclude in Section~\ref{sec:conclusion}.

\section{Vlasov equation}
\label{sec:Vlasov}

In this section we will introduce Vlasov equation and briefly review existing Vlasov 
solvers. Many more details can be found in 
Ref.~\cite{ref:Chen} in the context of plasma physics in general, and in 
Ref.~\cite[chap.~6]{ref:Chao} for the specific case of impedance effects in beam physics.

\subsection{Particle distribution in phase space}

\begin{figure}[ht]
\begin{center}
\includegraphics[width=0.5\textwidth]{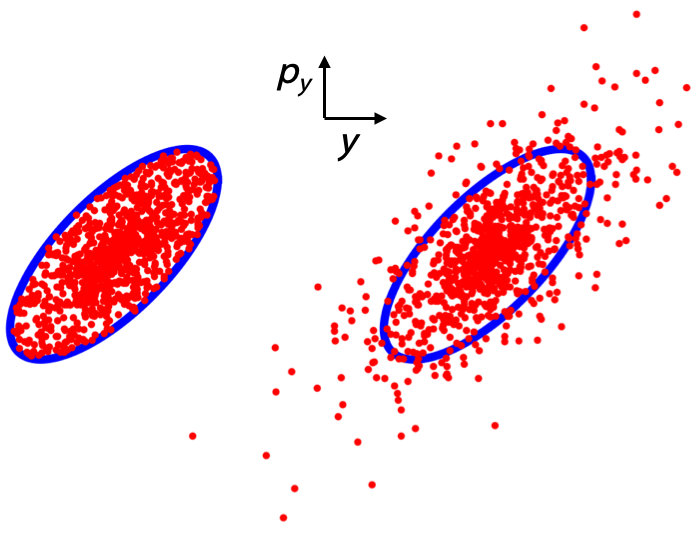}
\caption{Uniform (left) and Gaussian (right) distributions of particles in phase space, 
projected on the vertical plane $(y,p_y)$. The ellipse contains all the particles of the 
uniform distribution, and 80\% of them for the Gaussian case.}
\label{fig:distributions}
\end{center}
\end{figure}

In a classical picture (i.e. neglecting quantum-mechanical effects), each beam particle 
has a well defined position and momentum in phase space for each of the three 
coordinates $(x,y,z)$. The distribution function $\psi$ then represents the 
\textit{density} of particles in the six-dimensional (6D) phase space. 
In~\Fref{fig:distributions} we sketch examples of such particles distributions,
represented here in one plane only.

The number of particles in a given phase space volume $V$ is then obtained from the 6D integral
\begin{equation}
 \mathcal{N}(V)=\iiint\iiint_{V} \mathrm{d} x \mathrm{d} y \mathrm{d} z \,
\mathrm{d} p_{x} \mathrm{d} p_{y} \mathrm{d} p_{z} \; \psi\left(x, y, z, p_{x}, p_{y}, 
p_{z}\right),
\end{equation}
which gives the total number of particles $N$ when $V$ is the full phase 
space. The average value of any function $f$ of the particle positions and momenta, over 
a phase space volume $V$, is similarly given by
\begin{equation}
 \left<f\left(V\right)\right>=\iiint\iiint_{V} \mathrm{d} x \mathrm{d} y 
\mathrm{d} z \, 
\mathrm{d} p_{x} 
\mathrm{d} p_{y} \mathrm{d} p_{z} \; f\left(x, y, z, p_{x}, p_{y}, p_{z}\right) 
\psi\left(x, y, z, p_{x}, p_{y}, p_{z}\right).
\end{equation}

In the case of an ensemble of particles evolving with time, $\psi$ also depends on the time $t$. 

\subsection{From Liouville theorem to Vlasov equation}

\begin{figure}[ht]
\begin{center}
\includegraphics[width=0.7\textwidth]{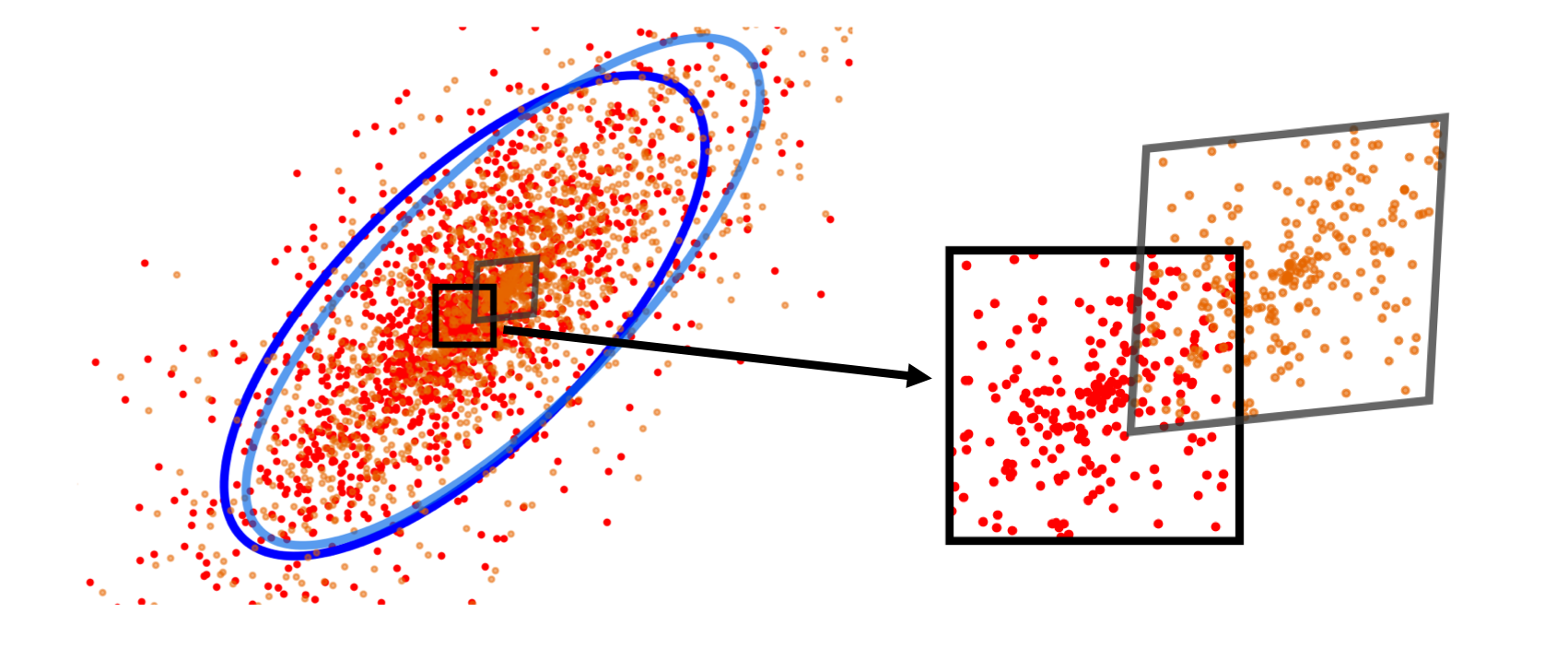}
\caption{Example of time evolution of a Gaussian distribution of particles. Red particles 
at time $t$ become the orange ones at time $t+d t$, and the associated dark blue ellipse 
(resp. the black square) becomes the light blue one (resp. the grey parallelogram).}
\label{fig:Liouville}
\end{center}
\end{figure}

In a collisionless Hamiltonian system, \textit{Liouville's theorem} states that the 
\textit{distribution function is constant along any trajectory in phase space}. This 
means 
that the local phase space distribution function does not change when one follows the 
flow 
(i.e. the trajectory) of particles. This is illustrated in~\Fref{fig:Liouville}: time 
evolution makes the square evolve into a parallelogram of same area, containing the same number of 
particles as the initial square, hence the average density will be the same.

Mathematically, this theorem can be expressed by stating that the \textit{total time 
derivative} of the distribution function is zero at any time, i.e.

\begin{equation}
 \frac{d \psi}{d t}=0 . \label{eq:Liouville}
\end{equation}

Liouville's theorem applies in principle to a system of non-colliding (i.e. 
non-interacting) particles -- it is actually equivalent to the \textit{collisionless} 
Boltzmann transport equation. For a plasma such as a beam of particles in a synchrotron, 
one could think of the particle electromagnetic (EM) interactions as Coulomb collisions 
in 
the rest frame of the beam, hence rather intuitively use a \textit{collision-based} 
Boltzmann equation, with Coulomb interactions added as a collision term, rather than 
Liouville's theorem. In the early days of plasma physics, this approach 
turned out not to be successful, essentially because of the long-range character of 
electromagnetic interactions. A much better approach, found by 
Vlasov~\cite{ref:Vlasov_in_Russian,ref:Vlasov}, 
was instead to integrate the collective, self-interaction EM fields 
into 
the Hamiltonian itself. This gave birth to \textit{Vlasov equation} which is simply the 
expression of~\Eref{eq:Liouville} for a plasma under the action of its own EM fields.

The only assumptions for the Vlasov equation are therefore:
\begin{itemize}
 \item the system is Hamiltonian; in particular, there is no damping or diffusion 
mechanism due to external sources\footnote{Synchrotron damping is therefore 
neglected here - when this effect is significant (e.g. in electron synchrotrons) one 
would rather need to solve a Fokker-Planck 
equation~\cite{ref:Chen,ref:Chandrasekhar,ref:WarnockVlasovFokkerPlanck2006,ref:Lindberg}}
,
\item particles are interacting only through the collective EM fields (no short-range 
collisions),
\item there is no creation nor annihilation of particles.
\end{itemize}

\subsection{A short review of existing Vlasov solvers}

Vlasov solvers have been theorized and implemented since as early as 1965 with the 
pioneer work of Laslett et al~\cite{ref:Laslett}. They can be used for various kinds of 
collective effects involving self-generated EM fields:
\begin{itemize}
 \item beam-beam effects~\cite{ref:Chao1984,ref:Alexahin},
 \item electron-cloud (and more generally two-stream 
effects)~\cite{ref:Perevedentsev,ref:Davidson},
 \item space-charge\cite{ref:Ryne},
 \item combined effect of impedance and 
space-charge~\cite{ref:Blaskiewicz,ref:Burov,ref:Balbekov,ref:Shobuda},
 \item longitudinal 
impedance~\cite{ref:Venturini,ref:Lindberg_longitudinal},
 \item transverse 
impedance~\cite{ref:Laslett,ref:Sacherer1972,ref:Sacherer1974,ref:Zotter_EMajorana,ref:Sacherer_EMajorana,
ref:Besnier1979,ref:Besnier1984,ref:Chin_LandauDamping,ref:Laclare,ref:JSBerg,
ref:KarlinerPopov,ref:Lindberg, ref:Schenk}.
\end{itemize}

For impedance-related instabilities, a number of codes are available, implementing 
Vlasov solvers in longitudinal --- e.g. GALACLIC (GArnier-LAclare Coherent 
Longitudinal Instabilities 
Code)~\cite{ref:EMetral_GALACTIC_GALACLIC}, as well as in transverse --- e.g. MOSES 
(MOde-coupling Single bunch instability in an Electron Storage 
ring)~\cite{ref:MOSES1,ref:MOSES2}, NHTVS (Nested Head-Tail Vlasov 
Solver)~\cite{ref:NHTVS}, DELPHI (Discrete Expansion over Laguerre Polynomials and 
Head-tail modes to compute 
Instabilities)~\cite{ref:Mounet_Benevento}, and GALACTIC (GArnier-LAclare Coherent 
Transverse Instabilities
Code)~\cite{ref:EMetral_GALACTIC,ref:EMetral_GALACTIC_GALACLIC}.

\section{Linearized Vlasov equation}
\label{sec:linear}

After short introductions on Hamiltonians, canonical transformations and Poisson brackets, 
we 
will present in this section the basics of perturbation theory applied to Vlasov 
equation. A general linearized version of Vlasov equation will be derived, as done in 
Ref.~\cite{ref:KLi_Vlasov}. Higher order extensions are then also obtained.

\subsection{Hamiltonian formalism}

We consider the coordinates $(x,y,z)$ of a particle relative to a reference 
particle --- called the \textit{synchronous} particle --- which travels along the 
synchrotron orbit at 
exactly the design velocity $v=\beta c$, with $\beta$ the relativistic velocity factor 
and $c$ the speed of light, such that its longitudinal position along the orbit is 
$s=vt$. In this reference system the synchronous particle has therefore positions 
$(x=0,y=0,z=0)$. $z$ is tangential to the orbit, and $(x,y)$ perpendicular to $z$, 
with $x$ in the plane of the orbit and directed towards the outside. We adopt here 
essentially the conventions and most of the notations of Ref.~\cite[chap.~1]{ref:Chao}.

The beam particles are solely affected by electromagnetic forces and 
constitute therefore a Hamiltonian system. In general the Hamiltonian is defined as 
the Legendre transform of the Lagrangian, which itself is the difference between the 
kinetic energy and the potential energy of the system. We will state without proof that 
the Hamiltonian corresponds to the sum of the kinetic and potential energy, and 
refer the reader to Ref.~\cite{ref:Goldstein} for more details on classical Hamiltonian 
mechanics, and to Refs.~\cite{ref:Bell,ref:Ruth_Hamiltonian,ref:Herr} for a detailed 
description of Hamiltonian dynamics applied to single particle beam physics.

The single-particle dynamics will be here expressed thanks to an effective
Hamiltonian~\cite{ref:Herr} defined in general as a function of all the phase 
space coordinates and time
\begin{equation}
 H\left(x, x^{\prime}, y, y^{\prime}, z, \delta ; t\right),
\end{equation}
with
\begin{align}
  x^{\prime} \equiv \frac{d x}{d s}, \qquad 
  y^{\prime} \equiv \frac{d y}{d s}, \qquad 
  \delta \equiv \frac{\Delta p_z}{p_0}, 
\label{eq:transf_prime}
\end{align}
where $\Delta p_z$ is the deviation of the longitudinal momentum from that of the 
synchronous particle given by \[ p_0 = m_0 \gamma v,\] $m_0$ being the 
rest mass and \[ \gamma=\sqrt{\frac{1}{1-\beta^2}} \] the relativistic 
mass factor.

The time evolution is governed by Hamilton's equations. In general, for any conjugate
pair of position and momentum coordinates $(x_i,p_i)$, these are written
\begin{equation}
 \frac{d x_i}{d t}=\frac{\partial H}{\partial p_i}, \qquad \frac{d p_i}{d 
t}=-\frac{\partial H}{\partial x_i}. \label{eq:Hamilton_general}
\end{equation}

In our case, Hamilton's equations read
\begin{align}
 \frac{d x}{d t}&=\frac{\partial H}{\partial x^{\prime}}, \qquad \frac{d x^{\prime}}{d 
t}=-\frac{\partial H}{\partial x}, \nonumber \\
 \frac{d y}{d t}&=\frac{\partial H}{\partial y^{\prime}}, \qquad \frac{d y^{\prime}}{d 
t}=-\frac{\partial H}{\partial y}, \nonumber \\
 \frac{d z}{d t}&=\frac{\partial H}{\partial \delta}, \qquad \frac{d \delta}{d 
t}=-\frac{\partial H}{\partial z}. \label{eq:Hamilton}
\end{align}

\subsection{Canonical transformations and symplecticity}

A \textit{canonical transformation of coordinates} is a change of coordinates 
that preserves Hamilton's equations~\eqref{eq:Hamilton_general}. It can be 
proven~\cite{ref:Goldstein} that this condition is equivalent to the 
\textit{symplecticity} of the Jacobian $\mathcal{J}$ of the transformation of 
coordinates
\begin{equation}
\mathcal{J}^{T} \cdot S \cdot \mathcal{J}=S, \label{eq:symplecticity}
\end{equation}
where $(.)^T$ denotes the transpose of a matrix, while $S$ and $\mathcal{J}$ are given by
\begin{align}
S &= \begin{pmatrix}{0} & {\cdots} & {0} & {1} & {\cdots} & {0} \\ {\vdots} & 
{\ddots} & {\vdots} & {\vdots} & {\ddots} & {\vdots} \\ {0} & {\cdots} & {0} & {0} & 
{\cdots} & {1} \\ {-1} & {\cdots} & {0} & {0} & {\cdots} & {0} \\ {\vdots} & {\ddots} & 
{\vdots} & {\vdots} & {\ddots} & {\vdots} \\ {0} & {\cdots} & {-1} & {0} & {\cdots} & 
{0}\end{pmatrix}, \\ & \nonumber \\
 \mathcal{J} &= \begin{pmatrix}{\frac{\partial X_{1}}{\partial x_{1}}} & {\cdots} & 
{\frac{\partial X_{1}}{\partial x_{n}}} & {\frac{\partial X_{1}}{\partial p_{1}}} & 
{\cdots} & {\frac{\partial X_{1}}{\partial p_{n}}} \\ {\vdots} & {\ddots} & {\vdots} & 
{\vdots} & {\ddots} & {\vdots} \\ {\frac{\partial X_{n}}{\partial x_{1}}} & {\cdots} & 
{\frac{\partial X_{n}}{\partial x_{n}}} & {\frac{\partial X_{n}}{\partial p_{1}}} & 
{\cdots} & {\frac{\partial X_{n}}{\partial p_{n}}} \\ \\ 
{\frac{\partial P_{1}}{\partial x_{1}}} & {\cdots} & 
{\frac{\partial P_{1}}{\partial x_{n}}} & {\frac{\partial P_{1}}{\partial p_{1}}} & 
{\cdots} & {\frac{\partial P_{1}}{\partial p_{n}}} \\ {\vdots} & {\ddots} & {\vdots} & 
{\vdots} & {\ddots} & {\vdots} \\ {\frac{\partial P_{n}}{\partial x_{1}}} & {\cdots} & 
{\frac{\partial P_{n}}{\partial x_{n}}} & {\frac{\partial P_{n}}{\partial p_{1}}} & 
{\cdots} & {\frac{\partial P_{n}}{\partial p_{n}}}\end{pmatrix},
\end{align}
$\left(x_i\right)_{i=1..n}$ (resp. $\left(X_i\right)_{i=1..n}$) being the old (resp. 
new) positions, and $\left(p_i\right)_{i=1..n}$ (resp. $\left(P_i\right)_{i=1..n}$) the 
old (resp. new) momenta.
The condition of symplecticity in~\Eref{eq:symplecticity} also implies
\begin{equation}
 \det(\mathcal{J}) \det(S) \det(\mathcal{J}) = \det(S) \quad \Rightarrow \quad
\left|\det(\mathcal{J})\right| = 1,
\end{equation}
which means the phase space volume is conserved by a canonical 
transformation~\cite{ref:Bell}
\begin{equation}
\iint \prod^n_{i=1} \mathrm{d} X_{i} \mathrm{d} P_{i} = \iint \prod^n_{i=1} 
\mathrm{d} x_{i} \mathrm{d} p_{i}.
\end{equation}

\subsection{Poisson bracket}

For any positions $\left(x_i\right)_{i=1..n}$ and momenta $\left(p_i\right)_{i=1..n}$, the
Poisson bracket~\cite{ref:Goldstein} of two differentiable functions $f$ and $g$ is 
defined as
\begin{equation}
[f, g]=\sum_{i=1}^n \frac{\partial f}{\partial x_{i}} \frac{\partial g}{\partial 
p_{i}}-\frac{\partial f}{\partial p_{i}} \frac{\partial g}{\partial x_{i}}. 
\label{eq:Poisson}
\end{equation}

The Poisson bracket is a bilinear form (i.e. linear with respect to each of the functions 
$f$ and $g$). In a Hamiltonian system it can be used to express in a compact way the 
total derivative of any time dependent function $f\left(x_i,p_i;t\right)$ of the phase 
space coordinates
\begin{align}
  \frac{d f}{d t} &= \frac{\partial f}{\partial t} + \sum_{i=1}^n \frac{\partial 
f}{\partial x_i} \frac{d x_i}{d t} + \frac{\partial f}{\partial p_i} \frac{d 
p_i}{d t} \nonumber \\
  &= \frac{\partial f}{\partial t} + \sum_{i=1}^n \frac{\partial f}{\partial x_i}\ 
\frac{\partial H}{\partial p_i} -  \frac{\partial f}{\partial p_i} \frac{\partial 
H}{\partial x_i} \qquad \textrm{using Hamilton's equations~\eqref{eq:Hamilton_general}} \nonumber \\
  &= \frac{\partial f}{\partial t} + [f,H].
\end{align}

Applying the above to the phase space distribution function $\psi$, we can re-write 
Liouville's theorem from~\Eref{eq:Liouville} as

\begin{equation}
 \frac{d \psi}{d t} = \frac{\partial \psi}{\partial t} + [\psi,H] = 0.
\end{equation}

Finally, one fundamental property of the Poisson bracket is its \textit{invariance 
under any canonical transformation}~\cite{ref:Goldstein}.

\subsection{Linearized Vlasov equation}

Here we adopt the framework of perturbation theory and assume the Hamiltonian $H$ is the 
sum of two terms: an \textit{unperturbed} Hamiltonian $H_0$, and a 
\textit{first order perturbation} $\Delta H$
\begin{equation}
  H = H_0 + \Delta H.
\end{equation}

In the context of beam dynamics with collective effects, typically $H_0$ will be 
expressing the focusing of the particles around the design orbit in the transverse plane, 
and around the synchronous particle in longitudinal, while $\Delta H$ will correspond to
the collective effect under study.

For the unperturbed Hamiltonian $H_0$, Vlasov equation admits as stationary solution any 
distribution function $\psi_0$ such that
\begin{equation}
 \frac{\partial \psi_0}{\partial t} + [\psi_0,H_0] = 0. \label{eq:psi0_H0}
\end{equation}
In the case of a time-independent unperturbed Hamiltonian, this can be achieved by any 
$\psi_0$ that is a time-independent function of $H_0$ itself, or any time-independent 
function of the invariants of motion --- but we do not have to 
restrict ourselves to such specific cases. The only important aspect is that $\psi_0$ 
should be
known.

We can then consider as well a first order perturbation 
$\Delta \psi$ of the unperturbed distribution function $\psi_0$, such that the total 
distribution function $\psi$ reads
\begin{equation}
\psi=\psi_{0}+\Delta \psi.
\end{equation}

Using Poisson bracket bilinearity, Vlasov equation can then be written
\begin{align}
 & \frac{\partial \psi}{\partial t}+[\psi, H]=0 \nonumber \\ 
 \Leftrightarrow \qquad & \frac{\partial\left(\psi_{0}+\Delta \psi\right)}{\partial 
t}+\left[\psi_{0}+\Delta \psi, H_{0}+\Delta H\right]=0 \nonumber \\
 \Leftrightarrow \qquad & \left( \frac{\partial \psi_0}{\partial t} + [\psi_0,H_0]\right) 
+ \left( \frac{\partial \Delta \psi}{\partial t}+\left[\Delta \psi, 
H_{0}\right]+\left[\psi_{0}, \Delta H\right] \right) + [\Delta \psi, \Delta H] = 0.
\end{align}

In the final equation, the first term between brackets is zero by virtue 
of~\Eref{eq:psi0_H0}, while the last one is of second order so should be neglected in a 
first order perturbation theory. Remains therefore only the second term between brackets, 
and we get
\begin{equation}
 \frac{\partial \Delta \psi}{\partial t}+\left[\Delta \psi, 
H_{0}\right] = -\left[\psi_{0}, \Delta H\right]. \label{eq:linear_Vlasov}
\end{equation}

This is the \emph{linearized Vlasov equation}~\cite{ref:KLi_Vlasov}, in which the 
unknown is $\Delta \psi$. In this form, the equation is invariant under canonical 
transformations.

\subsection{Extension to higher orders}

As soon as the stationary distribution and its first order perturbation are known, we 
can in principle try to compute the next order. Writing now the total Hamiltonian and 
distribution function as
\begin{align}
 H &= H_0 + \Delta H + \Delta^2 H, \\
 \psi &= \psi_0 + \Delta \psi + \Delta^2 \psi,
\end{align}
Vlasov equation becomes
\begin{align}
 & \frac{\partial \psi}{\partial t}+[\psi, H]=0 \nonumber \\ 
 \Leftrightarrow \qquad & \frac{\partial\left(\psi_{0}+\Delta \psi+\Delta^2 
\psi\right)}{\partial 
t}+\left[\psi_{0}+\Delta \psi+\Delta^2 
\psi, H_{0}+\Delta H+\Delta^2 H\right]=0 \nonumber \\
 \Leftrightarrow \qquad & \left( \frac{\partial \psi_0}{\partial t} + [\psi_0,H_0]\right) 
+ \left( \frac{\partial \Delta \psi}{\partial t}+\left[\Delta \psi, 
H_{0}\right]+\left[\psi_{0}, \Delta H\right] \right) + \nonumber \\
& \qquad \qquad \qquad \left(\frac{\partial 
\Delta^2 \psi}{\partial t}+[\Delta \psi, \Delta H]+[\Delta^2 \psi, H_0]+[\psi_0, \Delta^2 
H]\right) + \; \textrm{higher-order terms} = 0.
\end{align}
$\psi_0$ cancels out the first term in brackets, and $\Delta\psi$ the second 
one, so we get at second order
\begin{equation}
  \frac{\partial \Delta^2 \psi}{\partial t}+[\Delta^2 \psi, 
H_0] = -[\Delta \psi, \Delta H]-[\psi_0, \Delta^2 H].
\end{equation}
This is the second order of Vlasov equation. It is worth noticing that the homogeneous 
part of the equation (the left-hand side) remains similar to that of the first order 
equation\footnote{Note, however, that for collective effects $\Delta 
H$ and $\Delta^2 H$ also depend on the perturbed distribution function.}. 
Also, the right-hand side is non-zero even if $H$ remains of first order (i.e. $ \Delta^2 
H = 0$).

Generalizing to any order with the notations
\begin{align}
  H &= \sum_{i=0}^n \Delta^i H, \nonumber \\
  \psi &= \sum_{i=0}^n \Delta^i \psi,
\end{align}
(such that $\Delta^0 H \equiv H_0$, $\Delta^1 H \equiv \Delta H$, etc.), we obtain
\begin{equation}
  \frac{\partial \Delta^n \psi}{\partial t}+\sum_{i=0}^n [\Delta^i \psi, 
\Delta^{n-i} H] = 0.
\end{equation}

\section{Building a Vlasov solver}
\label{sec:method}

Vlasov equation --- or its linearized form in~\Eref{eq:linear_Vlasov} --- is in 
principle a 
partial differential equation of seven variables (the six phase space coordinates and 
time). Solving it "brute force" would be not only intractable and 
overly time consuming with the currently available computer power, but also losing any 
asset with respect to other methods such as macroparticle tracking simulations, in 
particular concerning the physical understanding it can possibly provide.

As a consequence, building a useful Vlasov solver requires typically a significant 
analytical work in order to simplify the equation. The main idea is essentially to reduce 
its number of variables. In the case of instabilities, another core aspect is the 
identification of eigenmodes.

In this section we hence sketch an approach to perform such an analytical work. The main 
steps are described in details using a well-known example, the case of coherent 
transverse instabilities of a bunched beam, arising from the effect of dipolar impedance (i.e. the part of 
the transverse EM force between two particles, that is proportional to the transverse 
position of the particle creating the EM field --- see e.g. Ref.~\cite{ref:Zotter}). We 
will include the effect of chromaticity but not that of coupling between $x$ and $y$. At 
each step of the approach we will try to present its main general idea, as much as 
possible independently from the specific case we study, such that, in principle, the same 
outline could be followed for other kinds of Vlasov solvers. A number of assumptions and 
approximations will be made throughout this example, which are mainly taken from the 
derivation done in Ref.~\cite[chap.~6]{ref:Chao} and will be given sometimes without 
proving their applicability, as the point here is to describe a method rather than to 
provide a complete discussion on the validity of the theory.

The first of these assumptions is to assume a four-dimensional phase space 
$(y,y^{\prime},z,\delta)$, because the transverse motion is uncoupled (and we choose here 
the vertical plane). Hence the distribution function $\psi$ and the Hamiltonian $H$ will 
both depend only on those variables, and on time.

\subsection{Writing the Hamiltonian}
\label{sec:Hamiltonian}

Given the form of the linearized Vlasov equation~(\ref{eq:linear_Vlasov}), the first step 
is necessarily to write the Hamiltonian of the system.

We consider first the unperturbed part $H_0$. Rather than getting it from first principles,
we adopt the much simpler approach of writing an effective Hamiltonian starting from the
known equations of single particle dynamics. Single particles obey Hill's equation in 
vertical, and we assume that the 
smooth approximation holds~\cite[chap.~1]{ref:Chao}, i.e.
\begin{equation}
 \frac{d^2 y}{d s^2} + \left(\frac{Q_{y0}}{R}\right)^2 y = 0, \label{eq:Hill}
\end{equation}
with $Q_{y0}$ the unperturbed vertical tune of the circular machine, and $R$ its 
circumference divided by $2\pi$.
For the longitudinal plane $(z,\delta)$, we assume a linear RF bucket, and the equations 
of motion are~\cite[chap.~1]{ref:Chao}
\begin{align}
 z^{\prime} &= -\eta \delta \nonumber \\
 \delta^{\prime} &= \frac{1}{\eta} \left(\frac{Q_s}{R}\right)^2 z, \label{eq:RF}
\end{align}
with $Q_s$ the synchrotron tune and $\eta=\alpha_p-\frac{1}{\gamma^2}$ the slippage 
factor. $\alpha_p$ is a property of the machine optics, called \textit{momentum 
compaction factor} --- this is the proportionality factor between the relative change of 
orbit length and the relative momentum deviation of a particle.

To write the effective Hamiltonian $H_0$ we need to integrate back Hamilton's 
equations~(\ref{eq:Hamilton}). For the vertical plane these are
\begin{align}
 \frac{\partial H_0}{\partial y'} &= \frac{d y}{d t} = \frac{d y}{d s}\frac{ds}{dt} = v 
y', \nonumber \\
 \frac{\partial H_0}{\partial y} &= -\frac{d y'}{d t} = - v y'' = v 
\left(\frac{Q_{y0}}{R}\right)^2 y,
\end{align}
where in the very last step we have used Hill's equation~(\ref{eq:Hill}). We can 
integrate these to get the vertical dependence of the unperturbed Hamiltonian
\begin{equation}
 H_0 = \frac{v}{2} \left[\left(\frac{Q_{y0}}{R}\right)^2 y^2 + y'^2 \right] + 
\textrm{function of }(z,\delta) .
\end{equation}
The longitudinal dependence of $H_0$ is obtained in a similar way:
\begin{align}
 \frac{\partial H_0}{\partial \delta} &= \frac{d z}{d t} = v z' = - v \eta \delta, 
\label{eq:dH0_ddelta} \\
 \frac{\partial H_0}{\partial z} &= -\frac{d \delta}{d t} = - v \delta' = 
-\frac{v}{\eta}\left(\frac{Q_s}{R}\right)^2 z, \label{eq:dH0_dz}
\end{align}
using Eqs.~\eqref{eq:RF}. Upon integration these give
\begin{equation}
 H_0 = \frac{v}{2} \left[\left(\frac{Q_{y0}}{R}\right)^2 y^2 + y'^2 \right] - \frac{v}{2} 
\left[ \frac{1}{\eta} \left(\frac{Q_s}{R}\right)^2 z^2 + \eta \delta^2 \right] .
\end{equation}
The perturbation to the Hamiltonian coming from the collective effect under study, 
$\Delta H$, can also be found by integrating Hamilton's equations. To do so we 
use Newton's second law of motion to express the change of vertical momentum due to the 
coherent EM force $F^{coh}_y$ (that we do not need to specify at this stage) and we get
\begin{align}
 \frac{\partial \Delta H}{\partial y} &= - \left( \frac{d y'}{d t} 
\right)^{coh} = -\frac{1}{m_0 \gamma v} \left(\frac{d p_y}{dt}\right)^{coh} = 
-\frac{F_y^{coh}}{m_0 \gamma v},
\end{align}
with $p_y$ the vertical momentum. Hence, assuming 
$F_y^{coh}$ depends only on $z$ and on time $t$ (which is the case of dipolar impedance)
\begin{equation}
 \Delta H = - \frac{y F_y^{coh}\left(z;t\right)}{m_0 \gamma v}.
\end{equation}
The total Hamiltonian obtained here is consistent with 
Refs.~\cite{ref:Chao,ref:Mounet_Benevento}; the additional multiplicative factor of $v$ 
that is present here is consistent with the fact that
in those references the independent variable 
considered is $s\equiv vt$ rather than $t$ -- these are equivalent within our simple 
accelerator model. Note also that Ref.\cite[chap.~6]{ref:Chao} considers 
that $v=c$ ($\beta=1$), while here $\beta$ can be smaller than one.

\subsection{Choosing the right coordinates}
\label{sec:coord}

Since the Poisson brackets in the linearized Vlasov equation~(\ref{eq:linear_Vlasov}) are 
invariant under canonical transformations of the coordinates, we can choose the system of 
coordinates that makes the Hamiltonian as simple as possible. Quite naturally, the 
action-angle variables seem to be the most adequate to simplify the unperturbed 
Hamiltonian $H_0$:
\begin{align}
 J_{y} &=\frac{1}{2}\left(\frac{Q_{y 0}}{R} y^{2} + \frac{R}{Q_{y 0}} y^{\prime 2} 
\right), \quad & \theta_{y} &=\operatorname{atan}\left(\frac{R y^{\prime}}{Q_{y 0} 
y}\right), \label{eq:Jy_thetay} \\
 J_{z} &=\frac{1}{2}\left(\frac{\omega_{s}}{v \eta} z^{2}+\frac{v \eta}{\omega_{s}} 
\delta^{2}\right), \quad & \phi &=\operatorname{atan}\left(\frac{v \eta 
\delta}{\omega_{s} z}\right) .
\end{align}
The 'old' coordinates $(y,y',z,\delta)$ can be expressed as a function of the 'new' ones 
thanks to
\begin{align}
 y &=\sqrt{\frac{2 J_{y} R}{Q_{y0}}} \cos \theta_{y}, \quad & y^{\prime} &=\sqrt{\frac{2 
J_{y} Q_{y0}}{R}} \sin \theta_{y} ,\label{eq:y_yprime} \\
 z &=\sqrt{\frac{2 J_{z} v \eta}{\omega_{s}}} \cos \phi, \quad & \delta &=\sqrt{\frac{2 
J_{z} \omega_{s}}{v \eta}} \sin \phi.
\end{align}
We show in Appendix~\ref{app:symplectic} that this transformation of coordinates is 
indeed canonical. The unperturbed Hamiltonian becomes, in this new system of coordinates
\begin{equation}
 H_{0}=\omega_{0} Q_{y0} J_{y}-\omega_{s} J_{z}, \label{eq:H0_nochroma}
\end{equation}
with the angular frequency of revolution given by \[ 
\omega_0\equiv\frac{2\pi}{T_{rev}}=\frac{v}{R}, \] and the synchrotron angular
frequency \[\omega_s = \omega_0 Q_s .\] The 
perturbation to the Hamiltonian is then given by
\begin{equation}
 \Delta H=-\sqrt{\frac{2 J_{y} R}{Q_{y0}}} \cos \theta_{y} 
\frac{F_y^{coh}\left(z;t\right)}{m_{0} \gamma v} . \label{eq:DeltaH}
\end{equation}
In all the above, the tune $Q_{y0}$ is considered to be a constant. As in 
Ref.~\cite[chap.~6]{ref:Chao}, we now introduce a dependence of the tune on $\delta$ in 
the model, by replacing $Q_{y0}$ with \[ Q_y=Q_{y0}+Q'_y \delta,\] $Q'_y$ being 
the (unnormalized) \emph{chromaticity}. We get then
\begin{align}
 H_{0} &= \omega_{0} \left(Q_{y0}+Q'_y \delta\right) J_{y} - \omega_{s} J_{z} \nonumber \\
 &= \omega_0 Q_y J_y - \omega_s J_z. 
\label{eq:H0_chroma}
\end{align}
However, this change is done only in $H_0$ as expressed as a function of $J_y$ and $J_z$, 
and \textit{not} in the definition of $J_y$ and $\theta_y$ which remain as 
expressed in~\Eref{eq:Jy_thetay} (and $(y,y')$ remains as in~\Eref{eq:y_yprime}, 
hence $\Delta H$ does not change as well). This standard approximation can be \textit{a 
posteriori} 
understood by computing the additional term in the longitudinal motion 
coming from chromaticity, $\frac{\partial H_0^{\textrm{chroma}}}{\partial 
\delta}$ resulting from this modification, and comparing it to the term $-v \eta \delta$ 
in~\Eref{eq:dH0_ddelta}
\begin{equation}
\frac{\partial H_0^{\textrm{chroma}}}{\partial \delta} = v y^2\frac{Q_{y0} Q'_{y}}{R^2}.
\end{equation}
To get an idea of the order of magnitude of this term, we can write that
\[ <y^2> \sim \sigma_y^2 = \frac{\epsilon^N_y}{\beta\gamma}\frac{R}{Q_{y0}}, \]
with $\sigma_y$ the vertical beam size, $\epsilon^N_y$ the vertical normalized emittance, 
and using again the smooth approximation for the average beta function 
$<\beta_y>\sim\frac{R}{Q_{y0}}$. Then, taking an extreme case where $|Q'_y|~\sim Q_{y0}$, 
we get
\begin{equation}
 <y^2>\frac{Q_{y0} | Q'_{y} |}{R^2} \sim \frac{\epsilon^N_y}{\beta \gamma <\beta_y>}
\end{equation}
which has to be compared to $|\eta \delta|$. There are typically orders of magnitude 
between these two numbers: for the Large Hadron Collider (LHC), $|\eta \delta| \sim 
10^{-8}$ while $\frac{\epsilon^N_y}{\beta \gamma <\beta_y>} \sim ~ 10^{-10}$, and in a 
low-energy ring like the Proton Synchrotron Booster (PSB) the order of magnitude of these 
two numbers would be respectively $10^{-4}$ and $10^{-6}$. The  assumption still breaks 
down when the synchrotron is operating close or exactly at transition energy 
($\eta=0$). Instabilities occurring at transition are beyond the scope of this lecture.

In the following, we will therefore neglect this additional term $\frac{\partial 
H_0^{\textrm{chroma}}}{\partial \delta}$ in the longitudinal motion, i.e. we will assume 
that

\begin{equation}
 \frac{\partial}{\partial \delta} \left( \omega_0 Q_y J_y \right) \approx 0, \quad 
\textrm{i.e.} \quad \frac{\partial H_0}{\partial \delta} \approx -v \eta \delta = 
\frac{\partial}{\partial 
\delta} \left(-\omega_s J_z\right) .\label{eq:approx_chroma}
\end{equation}
This approximation simply means that we neglect any effect of chromaticity on the 
longitudinal motion itself. More generally, in the following the 
transverse plane is assumed to have a negligible effect on the longitudinal one, i.e. the 
longitudinal motion is treated as unperturbed.

\subsection{Finding the unperturbed stationary distribution}

Before writing the linearized Vlasov equation, we need first to have an expression for the
unperturbed distribution $\psi_0$. As already stated, any function of the 
invariants of motion is a solution of Vlasov equation for $H_0$. The very simple form of 
$H_0$ in~\Eref{eq:H0_chroma} gives us readily the invariants as $J_y$ and (approximately) 
$J_z$: using the Poisson bracket definition from~\Eref{eq:Poisson} and applying it to our 
system of coordinates $(J_y,\theta_y,J_z,\phi)$ we have for any differentiable functions 
$f$ and $g$
\begin{equation}
 [f,g] = \frac{\partial f}{\partial J_y} \frac{\partial g}{\partial \theta_y} - 
\frac{\partial f}{\partial \theta_y}
  \frac{\partial g}{\partial J_y} + \frac{\partial f}{\partial J_z} \frac{\partial 
g}{\partial \phi} - \frac{\partial f}
  {\partial \phi} \frac{\partial g}{\partial J_z} ,
\end{equation}
such that
\begin{equation}
 \frac{d J_y}{dt} = [J_y,H_0] = \frac{\partial H_0}{\partial \theta_y}  = 0 ,
 \end{equation}
 and
\begin{align}
 \frac{d J_z}{dt} = [J_z,H_0] = \frac{\partial H_0}{\partial \phi}  &= 
 \frac{\partial H_0}{\partial \delta} \frac{\partial \delta}{\partial \phi}  + 
\frac{\partial H_0}{\partial z}
  \frac{\partial z}{\partial \phi} \nonumber \\
  &=  \frac{\partial H_0}{\partial \delta} \frac{\omega_s z}{v\eta}  - \frac{\partial 
H_0}{\partial z}  \frac{v\eta\delta}{\omega_s} \nonumber \\ 
  & \approx -\omega_s z \delta + \omega_s z \delta \nonumber \\
  & \approx 0 , \label{eq:approx_dJzdt}
\end{align}
where we have used Eq.~\eqref{eq:dH0_dz} and the approximation of the previous Section 
from Eq.~\eqref{eq:approx_chroma}.

Using the same approximation, the two planes can be assumed to be separated in the 
unperturbed distribution function
\begin{equation}
 \psi_0 = f_0 \left(J_y\right) g_0 \left(J_z\right), \label{eq:psi0}
\end{equation}
with $f_0$ and $g_0$ two one-dimensional functions --- resp. the vertical and 
longitudinal distribution functions. The phase space integral of $\psi_0$ should give the 
total number of particles $N$, so we can normalize these two functions such that
\begin{align}
 \int_0^{+\infty} d J_y \, f_0 \left(J_y\right) = \frac{N}{2\pi} ,\nonumber \\
 \int_0^{+\infty} d J_z \, g_0 \left(J_z\right) = \frac{1}{2\pi} , \label{eq:norm_f0_g0}
\end{align}
(the $2\pi$ denominators coming from the integration over the angles).

\subsection{Expressing the linearized Vlasov equation}

Now we turn to the determination of $\Delta \psi$, which is the core of the problem we 
want to solve. To do so we need to express the linearized Vlasov 
equation~\eqref{eq:linear_Vlasov}. 
Thanks to the preliminary work we have done and in particular our choice 
of variables, the Poisson bracket involving $H_0$ takes the particularly simple form
\begin{align}
 [\Delta\psi, H_0] &\approx -\frac{\partial \Delta \psi}{\partial \theta_y} \frac{\partial 
H_0}{\partial J_y} - \frac{\partial \Delta \psi}{\partial \phi} \frac{\partial 
H_0}{\partial J_z} \nonumber \\
 &\approx -\omega_0 Q_y \frac{\partial \Delta \psi}{\partial \theta_y} + \omega_s 
\frac{\partial \Delta \psi}{\partial \phi}, \label{eq:dpsi_H0}
\end{align}
where we again used the approximation~\eqref{eq:approx_chroma} in the form 
$\frac{\partial H_0}{\partial \phi} \approx 0$ as found in \Eref{eq:approx_dJzdt}.

The other Poisson bracket also takes a quite simple form: from Eqs.~\eqref{eq:DeltaH} 
and~\eqref{eq:psi0} we get
\begin{align}
 [\psi_0,\Delta H] &= \frac{d f_0}{d J_y} g_0(J_z) \frac{\partial \Delta H}{\partial 
\theta_y} + f_0(J_y) \frac{d g_0}{d J_z} \frac{\partial \Delta H}{\partial \phi} 
\nonumber \\
 &= \frac{d f_0}{d J_y} g_0(J_z) \sqrt{\frac{2 J_{y} R}{Q_{y0}}} \sin \theta_{y} 
\frac{F_y^{coh}\left(z;t\right)}{m_{0} \gamma v} + f_0(J_y) \frac{d g_0}{d J_z} 
\frac{\partial \Delta H}{\partial z} \frac{\partial z}{\partial \phi} \nonumber\\
 &\approx \frac{d f_0}{d J_y} g_0(J_z) \sqrt{\frac{2 J_{y} R}{Q_{y0}}} \sin \theta_{y} 
\frac{F_y^{coh}\left(z;t\right)}{m_{0} \gamma v}. \label{eq:psi0_dH}
\end{align}
In the above we have neglected $\frac{\partial \Delta H}{\partial z}$. This is a standard 
approximation, also made in Ref.~\cite[chap.~6]{ref:Chao}, which has its grounds in the 
general idea that we neglect
any effect of the transverse coherent force on the longitudinal motion. This should be 
valid in the context of transverse impedance effects as long as one remains far from 
low-order synchro-betatron resonances $Q_{y0}+l Q_s=n$ (and provided the transverse beam size is 
small enough).

Combining Eqs.~\eqref{eq:linear_Vlasov},~\eqref{eq:dpsi_H0} and~\eqref{eq:psi0_dH} we 
get the linearized Vlasov equation as
\begin{equation}
 \frac{\partial \Delta \psi}{\partial t} - \omega_0 Q_y \frac{\partial \Delta 
\psi}{\partial \theta_y} + \omega_s 
\frac{\partial \Delta \psi}{\partial \phi}= - \frac{d f_0}{d J_y} g_0(J_z) \sqrt{\frac{2 
J_{y} R}{Q_{y0}}} \sin \theta_{y} 
\frac{F_y^{coh}\left(z;t\right)}{m_{0} \gamma v}. \label{eq:linear_Vlasov_specific}
\end{equation}

\subsection{Writing and decomposing the perturbation}

Up to now, we have kept a general, unspecified form of $\Delta \psi$. 
To reduce further the complexity of the problem, we now need to specify the form of 
$\Delta \psi$. We will do so by giving it a time dependence in the 
form of a complex exponential: since we are looking for unstable modes in the vertical 
plane, we can assume that there is a single oscillation mode with a certain complex 
angular frequency $\Omega$, close to $\omega_0 Q_{y0}$ (the latter being the natural
oscillation frequency of the vertical plane)
\begin{equation}
 \Delta \psi (J_y,\theta_y,J_z,\phi; t) = \Delta \psi (J_y,\theta_y,J_z,\phi) e^{j\Omega t} ,
\end{equation}
with $j$ the imaginary unit\footnote{Note the replacement of $-i$ from 
Ref.~\cite{ref:Chao} by $j$ (here we use indeed the $e^{j\omega t}$ convention for the 
Fourier transform).}.
The idea here is that the mode of frequency $\Omega$ is the strongest one, superseding 
exponentially any other mode.

We also make a new change of variable in the longitudinal 
plane, out of convenience, the idea being to simplify slightly the equations and to keep 
similar expressions as in Ref.~\cite[chap.~6]{ref:Chao}:
\begin{equation}
 r=\sqrt{\frac{2 J_{z} v \eta}{\omega_{s}}}, \quad z=r \cos \phi, \quad \frac{v 
\eta}{\omega_{s}} \delta=r \sin \phi .
\end{equation}
Note that at this stage, we are not using anymore Hamilton's equations nor Poisson 
brackets with these new variables, so the transformation does not have to be canonical.

Even with the time dependence of $\Delta \psi$ thus 
specified,~\Eref{eq:linear_Vlasov_specific} above is still a partial 
differential equation involving four variables (and two different partial derivatives of 
the unknown $\Delta \psi$). It is quite natural to simplify the equation by 
performing a Fourier series decomposition of each angle $\theta_y$ and $\phi$, in order 
to be able to perform term-by-term identification in the resulting Vlasov equation. We 
do it first with $\theta_y$:
\begin{equation}
 \Delta \psi\left(J_{y}, \theta_{y}, r, \phi ; t\right)=e^{j \Omega t} 
\sum_{p=-\infty}^{+\infty} f^{p}\left(J_{y}\right) e^{j p \theta_{y}} g^p(r,\phi) .
\end{equation}
Such Fourier series expansion is exact, but in the above equation we have assumed that the dependencies 
over $J_y$ and over $(r,\phi)$ can be separated in each term in the series --- essentially we remain 
uncoupled between the vertical and the longitudinal planes.

Then we do a similar Fourier decomposition along the $\phi$ angle, but after extracting
away the headtail phase factor $e^{-\frac{j p Q_{y}^{\prime} z}{\eta R}}$ from 
$g^p(r,\phi)$, as done in Ref.~\cite[chap.~6]{ref:Chao} --- we can do it without loss of 
generality (this is just writing the Fourier series of $g^p(r,\phi) e^{\frac{j p 
Q_{y}^{\prime} z}{\eta R}}$ instead of $g^p(r,\phi)$ directly) and this will 
simplify significantly the equations in the next section:
\begin{equation}
 g^p(r,\phi) = e^{-\frac{j p Q_{y}^{\prime} z}{\eta R}} \cdot \sum_{l=-\infty}^{+\infty} 
R^p_{l}(r) e^{-j l \phi}. \label{eq:gp_r_phi}
\end{equation}
In the end, the perturbation to the distribution function is written as
\begin{equation}
\Delta \psi\left(J_{y}, \theta_{y}, r, \phi ; t\right) = e^{j \Omega t} 
\sum_{p=-\infty}^{+\infty} f^{p}\left(J_{y}\right) e^{j p \theta_{y}} \cdot e^{-\frac{j p 
Q_{y}^{\prime} z}{\eta R}} \cdot \sum_{l=-\infty}^{+\infty} R^p_{l}(r) e^{-j l \phi} . 
\label{eq:dpsi_decomp}
\end{equation}
The functions $f^p(J_y)$ and $R_l^p(r)$ are the ones that remain to be found.

\subsection{Reducing the number of variables}

With this decomposition of the perturbation, we will be able to significantly reduce the 
number of variables. First we express the needed partial derivatives of $\Delta \psi$ as
\begin{align}
 \frac{\partial \Delta \psi}{\partial t} &= j\Omega \Delta \psi = \left( e^{j 
\Omega t} \sum_{p=-\infty}^{+\infty} f^{p}\left(J_{y}\right) e^{j p \theta_{y}} \cdot 
e^{-\frac{j p Q_{y}^{\prime} z}{\eta R}} \cdot \sum_{l=-\infty}^{+\infty} R^p_{l}(r) 
e^{-j l \phi} \right) \times \left(j\Omega \right), \\
  \frac{\partial \Delta \psi}{\partial \theta_y} &= e^{j \Omega t} 
\sum_{p=-\infty}^{+\infty} \left( f^{p}\left(J_{y}\right) e^{j p \theta_{y}} \cdot 
e^{-\frac{j p Q_{y}^{\prime} z}{\eta R}} \cdot \sum_{l=-\infty}^{+\infty} R^p_{l}(r) 
e^{-j l \phi} \right) \times \left( j p \right), \\
 \frac{\partial \Delta \psi}{\partial \phi} &= e^{j \Omega t} \sum_{p=-\infty}^{+\infty} 
f^{p}\left(J_{y}\right) e^{j p \theta_{y}} \cdot e^{-\frac{j p 
Q_{y}^{\prime} z}{\eta R}} \cdot \sum_{l=-\infty}^{+\infty} R^p_{l}(r) e^{-j l \phi} 
\times \left(-j l + j p \frac{Q'_y}{Q_s} \delta \right),
\end{align}
using \[ \frac{\partial z}{\partial \phi} = - r \sin \phi = -\frac{v \eta 
\delta}{\omega_s} = -\frac{R \eta \delta}{Q_s} .\]
Then, plugging the three equations above into~\Eref{eq:linear_Vlasov_specific}, we 
notice that the term $-j p \omega_0 Q'_y\delta$ in $-\omega_0 Q_y \frac{\partial \Delta 
\psi}{\partial \theta_y}$, cancels out with the term $ j p \omega_s 
\frac{Q'_y}{Q_s}\delta$ in $\omega_s \frac{\partial \Delta \psi}{\partial \phi}$ --- 
this was precisely the point of writing $g^p(r,\phi)$ with the headtail phase factor in 
front of the Fourier series in~\Eref{eq:gp_r_phi}. In the end, using
\[ \sin \theta_y = \frac{e^{j\theta_y} - e^{-j\theta_y} }{2j}, \] we get
\begin{multline}
e^{j \Omega t} \sum_{p=-\infty}^{+\infty} f^{p}\left(J_{y}\right) e^{j p \theta_{y}} 
e^{-\frac{j p Q_{y}^{\prime} z}{\eta R}} \sum_{l=-\infty}^{+\infty} R^p_{l}(r) e^{-j l 
\phi}\left(j \Omega-j p \omega_{0} Q_{y 0} -j l \omega_{s}\right) \\
= -\frac{d f_{0}}{d J_y} g_{0}(r) \sqrt{\frac{2 J_{y} R}{Q_{y0}}} \left( \frac{e^{j 
\theta_{y}}-e^{-j \theta_{y}}}{2 j} \right) \frac{F_{y}^{coh}(z;t)}{m_{0} \gamma v}. 
\label{eq:linear_Vlasov_expanded}
\end{multline}
Since this is valid for any $\theta_y$ and $F_y^{coh}$ does not depend on 
$\theta_y$ (in the case of dipolar impedance), term-by-term identification leads to
\begin{equation}
 f^p\left(J_y \right) = 0 \; \textrm{for any } \, p \neq \pm 1
\end{equation}
Moreover, $\Omega$ is assumed to be close to $\omega_0 Q_{y0}$ such that
\[ \left| j \Omega - j\omega_{0} Q_{y 0} - jl \omega_{s} \right| << \left| j \Omega + 
j\omega_{0} Q_{y 0} - j l \omega_{s} \right|. \]
This means that the factor in front of $f^{-1}(J_y)$ in 
the left-hand side of~\Eref{eq:linear_Vlasov_expanded}, is significantly larger in 
absolute value, than that in front of $f^{1}(J_y)$. Since the two corresponding terms in 
the right-hand side are equal, we can assume that~\cite[chap.~6]{ref:Chao}
\begin{equation*}
 f^{-1}(J_y) << f^1(J_y) \quad \textrm{for any } J_y,
\end{equation*}
and hence we will neglect $f^{-1}(J_y)$:
\begin{equation}
 f^{-1}(J_y) \approx 0.
\end{equation}
Equation~\eqref{eq:linear_Vlasov_expanded} is then reduced to its $p=1$ term, which gives 
after simplification
\begin{equation}
 e^{j \Omega t} f\left(J_{y}\right) e^{\frac{-j Q_{y}^{\prime} 
z}{\eta R}} \sum_{l=-\infty}^{+\infty} R_{l}(r) e^{-j l \phi}\left(\Omega-Q_{y 0} 
\omega_{0}-l \omega_{s}\right) = \frac{d f_{0}}{d J_y} g_{0}(r) 
\sqrt{\frac{2 J_{y} R}{Q_{y0}}} \frac{F_{y}^{coh}(z ; t)}{2 m_{0} \gamma v} ,
\end{equation}
where we took away all superscripts ``$1$'' for more readability. The above equation is 
equivalent to
\begin{equation}
\sum_{l=-\infty}^{+\infty} R_{l}(r) e^{-j l 
\phi}\left[\frac{f\left(J_{y}\right)\left(\Omega-Q_{y 0} \omega_{0}-l 
\omega_{s}\right)}{\frac{d f_{0}}{d J_y} \sqrt{\frac{2 J_{y} R}{Q_{y0}}}}\right] = e^{-j 
\Omega t} e^{\frac{j Q_{y}^{\prime} z}{\eta R}} g_{0}(r) \frac{F_{y}^{coh}(z ; t)}{2 
m_{0} 
\gamma v} .
\end{equation}
The right-hand side of the above \textit{does not depend on} $J_y$, therefore the term 
between square brackets must be a constant. As a consequence
\begin{equation}
 f(J_y) \propto \frac{d f_{0}}{d J_y} \sqrt{\frac{2 J_{y} R}{Q_{y0}}} .
\end{equation}
Putting the proportionality constant into the $R_l(r)$, the perturbation of the 
distribution will now look like
\begin{equation}
 \Delta \psi(J_y,\theta_y, r, \phi; t) = e^{j \Omega t} e^{j \theta_{y}} \frac{d 
f_{0}}{d J_y} \sqrt{\frac{2 J_{y} R}{ Q_{y0}}} \cdot e^{-\frac{j Q_{y}^{\prime} z}{\eta 
R}} \cdot \sum_{l=-\infty}^{+\infty} R_{l}(r) e^{-j l \phi}. \label{eq:dpsi}
\end{equation}
Hence, we have fully specified the transverse part of the distribution function, and the 
only remaining unknowns are the \textit{one-dimensional functions} $R_l(r)$, which are 
solutions of the equation
\begin{equation}
 \sum_{l=-\infty}^{+\infty} R_{l}(r) e^{-j l \phi} \left(\Omega-Q_{y 0} \omega_{0}-l 
\omega_{s}\right) = e^{-j \Omega t} e^{\frac{j Q_{y}^{\prime} z}{\eta R}} g_{0}(r) 
\frac{F_{y}^{coh}(z ; t)}{2 m_{0} \gamma v} . \label{eq:linear_Vlasov_Rl}
\end{equation}

\subsection{Writing the coherent electromagnetic force}

Up to now, the coherent force from the collective effect under study, has remained 
largely unspecified, which allowed us to keep the equations relatively compact. We merely
had to assume that $F_y^{coh}$ should depend only on the longitudinal position and on 
time (which, however, already excludes from our treatment a number of collective
forces, such that those arising from an electron cloud, from beam-beam effects, or from
quadrupolar impedance). To go further, we need now to express the force as a function of the phase space 
distribution.

\begin{figure}[ht]
\begin{center}
\includegraphics[width=0.49\textwidth]{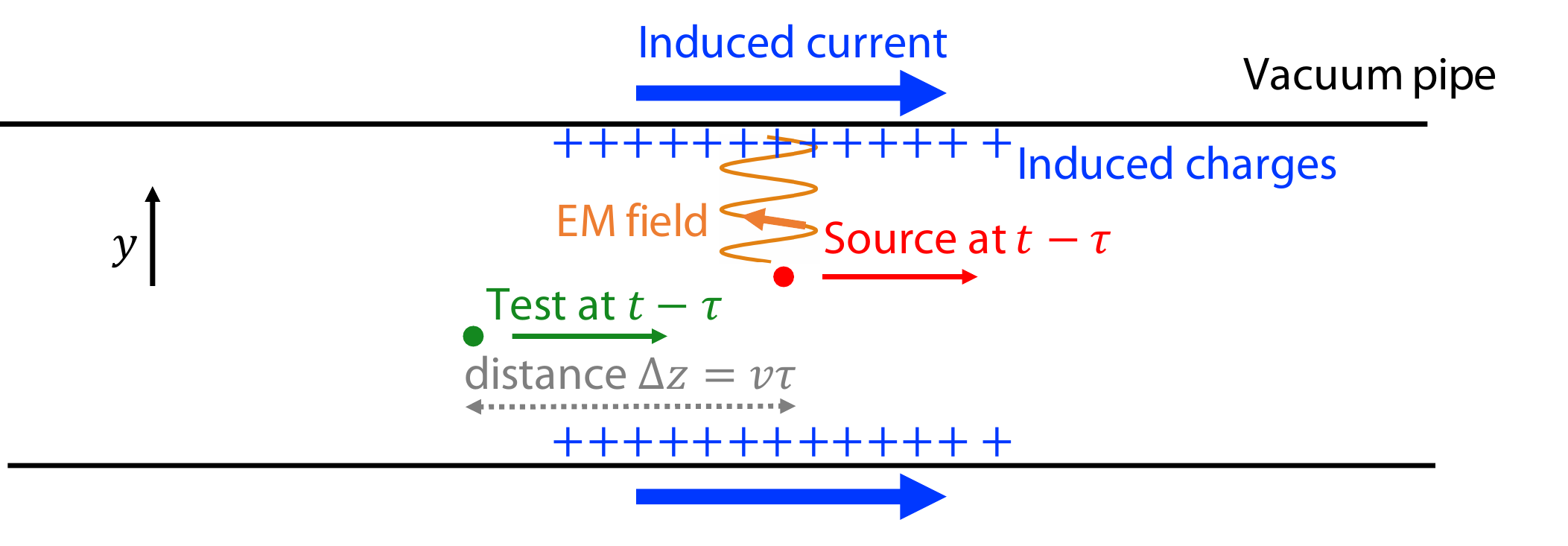}
\includegraphics[width=0.49\textwidth]{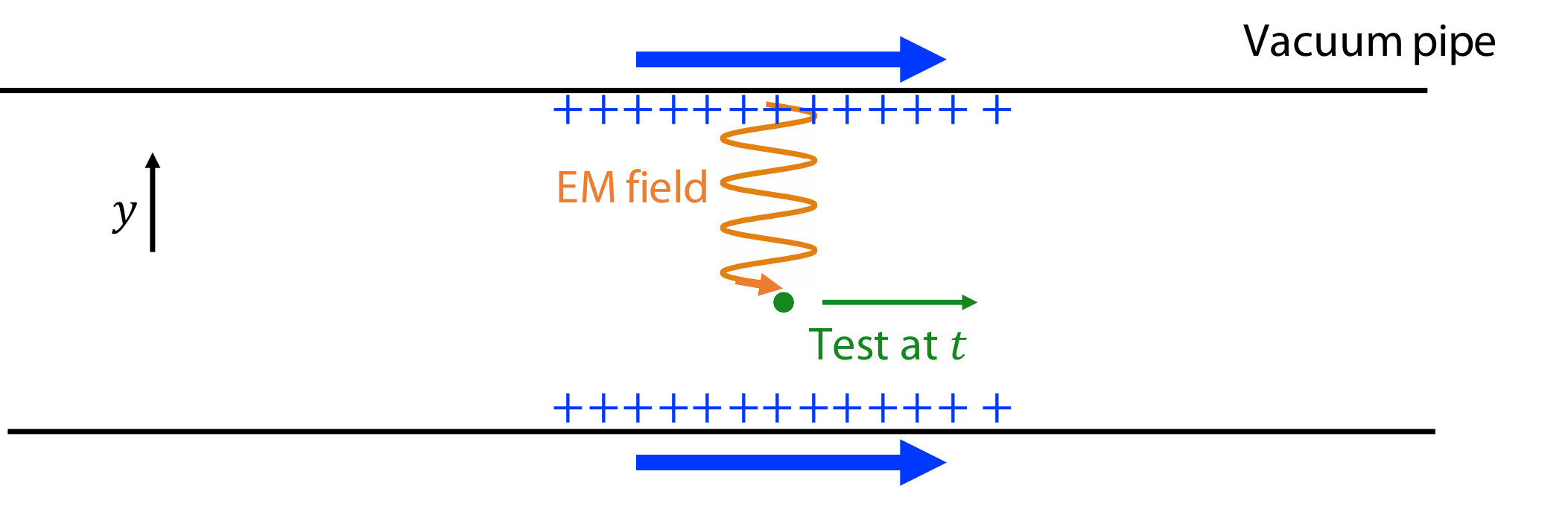} \hfill
\caption{Schematic representation of the impedance force}
\label{fig:impedance}
\end{center}
\end{figure}

In general, the impedance (or equivalently, the wake function) is in essence the EM force from a source 
particle, acting on a test particle typically located behind it (but not necessarily, in 
the case of non-ultrarelativistic machines). A schematic view of the impedance interaction 
is 
sketched in~\Fref{fig:impedance}. When a particle passes at time $t-\tau$ through an element 
in 
the machine (such as the vacuum pipe as in the plot, but it can also be a cavity, a 
kicker, a collimator, etc.), it creates an EM field that will interact with the 
surroundings, create there some induced charges and currents, and act back on a test 
particle coming at time $t$. The force between the two particles separated by a 
distance $\Delta z$, integrated over the length of the device $L$ and normalized by the 
particle charges, is called the wake function~\cite{ref:Bane_wake} and written
\begin{equation}
W_{y} \left(y_{\text{source}} , y_{\text{test}} , \Delta z\right) \equiv \frac{1}{e^{2}} \int_L 
ds F_{y} \left(y_{\text{source}} , y_{\text{test}} , \Delta z\right) = \frac{L}{e^2} \left< F_y 
\left(y_{\text{source}} , y_{\text{test}} , \Delta z\right) \right>, \label{eq:wake}
\end{equation}
where $\left< \cdot \right>$ represents the average over the length.
The force depends in principle on both the source and test vertical positions 
$y_{\text{source}}$ and $y_{\text{test}}$. In the case of the transverse dipolar 
impedance, which is the main source of coherent instabilities and hence the subject of 
this lecture, only source particles with an offset transverse position will create a force,
proportional to this offset and acting on the test particle independently from its 
position. Hence, applying ~\Eref{eq:wake} to the dipolar wake case, inverting it, and 
taking $L=2\pi R$ for the full wake function model of a synchrotron, we get
\begin{equation}
\left< F_y \right> \left(y_{\text{source}}, \Delta z\right)  = \frac{e^2}{2\pi R}  y_{\text{source}}
W^{dip}_{y} \left(\Delta z\right) . \label{eq:force_dip}
\end{equation}
Now $W^{dip}_{y}$ represents the proportionality factor between the total wake function $W_y 
\left(y_{\text{source}} , y_{\text{test}} , \Delta z\right)$ and the vertical offset of 
the source $y_{\text{source}}$. The average represents the integrated EM force over the 
full circumference~\cite[p. 57]{ref:Chao}. Indeed, the exact time (or $s$) dependence of the force
on the time scale of a single turn does not matter, as long as the rise time of the 
coherent oscillations is much longer than the revolution time; in essence we are 
considering a \textit{lumped} impedance acting on the beam at a single location around the ring.

Then, to compute the total force on a particle at a given 
position $(y,z)$ \textit{from the ensemble of particles}, we need to integrate the force 
over the phase space, using the total distribution function $\psi$
\begin{equation}
F^{coh}_{y}(z;t) = \frac{e^{2}}{2 \pi R} \iint d\tilde{z} d \tilde{\delta} \iint d \tilde{y} d \tilde{y}^{\prime} 
\psi\left(\tilde{y}, \tilde{y}^{\prime}, \tilde{z}, \tilde{\delta} ; t \right) \tilde{y} W^{dip}_{y}(\tilde{z}-z) . 
\label{eq:force_nomultiturn}
\end{equation}
Here $\tilde{y}$ inside the integral represents the source vertical coordinate; the 
force itself $F^{coh}_y(z)$ does not depend on the test vertical position $y$ since the wake is
purely dipolar.

In the expression above, we have not considered the fact that EM fields may remain at the 
location of the device while the beam is travelling several turns; in principle, one 
should therefore consider also sources from one turn before, two turns before, etc. (i.e. 
going through the impedance device at times $\frac{2\pi R}{v}$, $\frac{4\pi R}{v}$, $\frac{6\pi 
R}{v}$, etc.). The wake to be considered at each of these times has to be calculated for $\Delta 
z$ longer by the circumference times the number of turns of delay considered. Hence, we 
replace~\Eref{eq:force_nomultiturn} by
\begin{equation}
 F_{y}^{coh}(z; t)=\frac{e^{2}}{2 \pi R} \sum_{k=-\infty}^{+\infty} \iint d \tilde{z} d 
\tilde{\delta} \iint d \tilde{y} d \tilde{y}^{\prime} \psi\left(\tilde{y}, \tilde{y}^{\prime}, \tilde{z}, \tilde{\delta} ; t 
-k\frac{2\pi R}{v}\right) \tilde{y} W^{dip}_{y}(\tilde{z}+2 \pi k R - z). 
\label{eq:force_multiturn}
\end{equation}
Note that the sum extends not only to previous turns but also to \textit{forthcoming} 
turns ($k<0$), which is \textit{not} inconsistent with the causality principle: a 
particle passing at a given time $t$ will excite EM fields that 
can always catch back, at a later time, a non-ultrarelativistic test particle located in 
front of the source at any arbitrary distance $|\Delta z|$, even if $|\Delta z|$ reaches 
several full circumferences (provided we wait long 
enough). Then, the \textit{effect} of the force is actually computed 
not at this later time, but at $t$ (just like for the other $k\ge 0$ terms), considering 
the source was at $t+|\Delta z|/v$ and knowing the time evolution of the mode. 
In practice, even in low-energy machines the wake in 
front gets very rapidly extremely small, so the terms with $k<0$ are negligible anyway in 
most cases. Still, they are needed for the application of the Dirac comb formula (see below).

We further simplify~\Eref{eq:force_multiturn} by realizing that the unperturbed 
distribution $\psi_0$ is vertically centred, such that
\begin{equation}
 \iint d \tilde{y} d \tilde{y}^{\prime} \psi_{0}\left(\tilde{y}, \tilde{y}^{\prime}\right) \tilde{y}=0.
\end{equation}
(This can be seen from the fact that $\psi_0$ depends transversely only on $J_y$ and not on 
$\theta_y$, as visible in~\Eref{eq:psi0}, and by writing the above integral in 
action-angle coordinates).

Going to the $(J_y, \theta_y, r, \phi)$ system of coordinates, using~\Eref{eq:dpsi} and
\begin{align}
\iint d \tilde{z} d \tilde{\delta} &= \iint d \tilde{J_{z}} d \tilde{\phi}=\frac{\omega_{s}}{v \eta} 
\iint \tilde{r} d \tilde{r} d \tilde{\phi}, \\
\iint d \tilde{J_{y}} d \tilde{\theta_{y}} &= \iint d \tilde{y} d \tilde{y}^{\prime},
\end{align}
we get for the total force
\begin{multline}
F_{y}^{coh}(z;t)=\frac{e^{2}\omega_s}{2 \pi v \eta R} \sum_{k=-\infty}^{+\infty} e^{j 
\Omega \left(t-k \frac{2 \pi R}{v}\right)} \iint \tilde{r} d \tilde{r} d \tilde{\phi} \, W^{dip}_{y}(\tilde{r} \cos \tilde{\phi} 
+ 2 \pi k R-z) \\
\times \iint d \tilde{J_{y}} d \tilde{\theta_{y}} \, e^{j \tilde{\theta_{y}}} \frac{d 
f_{0}}{d \tilde{J_y}} \sqrt{\frac{2 \tilde{J_{y}} R}{ Q_{y0}}} \cdot e^{-\frac{j Q_{y}^{\prime} \tilde{r} 
\cos \tilde{\phi}}{\eta R}} \cdot \sum_{l=-\infty}^{+\infty} R_{l}(\tilde{r}) e^{-j l \tilde{\phi}} \sqrt{\frac{2 
\tilde{J_{y}} R}{Q_{y0}}} \cos \tilde{\theta_{y}} . \label{eq:force_action_angle}
\end{multline}
The multiturn sum \[ \sum_{k=-\infty}^{+\infty} e^{-j\Omega k \frac{2 \pi R}{v}} 
W^{dip}_{y}(\tilde{r} \cos \tilde{\phi} + 2 \pi k R-z),\] can be usefully transformed thanks to the 
frequency-domain counterpart of the wake function, i.e. the \textit{impedance}
$Z_y(\omega)$~\cite[chap. 2]{ref:Chao}
\begin{equation}
 W_{y}(z)=-\frac{j}{2 \pi} \int_{-\infty}^{+\infty} d \omega e^{j \omega \frac{z}{v}} 
Z_{y}(\omega),
\end{equation}
and the Dirac comb formula
\begin{equation}
 \sum_{k=-\infty}^{+\infty} e^{-j 2 \pi k \xi} = 
\sum_{k=-\infty}^{+\infty} \delta_D(\xi+k) \quad \text{for any} \; \xi,
\end{equation}
with $\delta_D$ the Dirac delta function. We obtain
\begin{align}
\sum_{k=-\infty}^{+\infty} e^{\frac{-j 2 \pi k \Omega R}{v}} W_{y}(\tilde{r} \cos \tilde{\phi} &+ 2 \pi k 
R-z) \nonumber \\
  &=\frac{-j}{2 \pi} \int_{-\infty}^{+\infty} d \omega Z_{y}(\omega) 
e^{j\omega\frac{\tilde{r}\cos\tilde{\phi} -z}{v}} \sum_{k=-\infty}^{+\infty} e^{-j 2 \pi k 
\frac{\Omega-\omega}{\omega_0}} \qquad \left( \text{with } \, \omega_0 = \frac{v}{R} 
\right) \nonumber \\
&=\frac{-j}{2 \pi} \int_{-\infty}^{+\infty} d \omega 
Z_{y}(\omega) e^{j\omega\frac{\tilde{r}\cos\tilde{\phi} -z}{v}} \sum_{k=-\infty}^{+\infty} \delta_D 
\left(\frac{\Omega-\omega}{\omega_0}+k\right) \nonumber \\
&=\frac{-j}{2 \pi} \int_{-\infty}^{+\infty} d \omega Z_{y}(\omega) 
e^{j\omega\frac{\tilde{r}\cos\tilde{\phi} -z}{v}} \sum_{k=-\infty}^{+\infty} \omega_0 \delta_D 
\left[\omega - \left(\Omega + k \omega_0\right) \right] \nonumber \\
&=\frac{-j \omega_{0}}{2 \pi} \sum_{k=-\infty}^{+\infty} 
Z_{y}\left(\Omega+k \omega_{0}\right) e^{j \left(\Omega+k\omega_0\right)\frac{\tilde{r}\cos\tilde{\phi} 
-z}{v}}.
\end{align}
Plugging this identity back into~\Eref{eq:force_action_angle} and re-ordering, we get
\begin{multline}
F_{y}^{coh}(z;t)= -e^{j \Omega t} \frac{j e^{2}\omega_s \omega_0}{2\pi^2 v \eta Q_{y0}} 
 \sum_{l=-\infty}^{+\infty} \sum_{k=-\infty}^{+\infty} Z_{y}\left(\Omega+k 
\omega_{0}\right) e^{-j \frac{\left(\Omega+k\omega_0\right) z}{v}} \int_0^{+\infty} \tilde{r} d \tilde{r} 
R_{l}(\tilde{r}) \\
\times \int_0^{2\pi} d \tilde{\phi} \cdot  e^{-j l \tilde{\phi}} e^{\frac{j \tilde{r}\cos\tilde{\phi}}{v} 
\left(\Omega+k\omega_0 - \frac{Q_{y}^{\prime}\omega_0}{\eta} \right)} \cdot \int_0^{+\infty} 
d \tilde{J_{y}} \frac{d f_{0}}{d \tilde{J_y}} \tilde{J_y} \cdot \int_0^{2\pi} d \tilde{\theta_{y}}
e^{j \tilde{\theta_{y}}} \cos \tilde{\theta_{y}} . \label{eq:force_impedance}
\end{multline}
The three latter integrals can be computed analytically, from
\begin{align}
 &\int_{0}^{2 \pi} d \tilde{\theta_{y}} \mathrm{e}^{j \tilde{\theta_{y}}} \cos \tilde{\theta_{y}} = 
\pi, \\
 &\int_{0}^{+\infty} d \tilde{J_{y}}  \; \tilde{J_{y}} \frac{d f_0}{d \tilde{J_y}} = \left[\tilde{J_{y}} 
f_{0}\left(\tilde{J_{y}}\right)\right]_{0}^{+\infty}-\int_{0}^{+\infty} d \tilde{J_{y}} 
f_{0}\left(\tilde{J_{y}}\right) = -\frac{N}{2 \pi},\\
 &\int_{0}^{2 \pi} d \tilde{\phi} e^{-j l \tilde{\phi}} e^{j \xi \cos \tilde{\phi}} = 2 \pi j^{l} 
J_{l}\left(\xi\right) \quad \text{ for any } \xi, \label{eq:Bessel}
\end{align}
using the normalization 
condition of the unperturbed distribution from~\Eref{eq:norm_f0_g0}, and Eq.~(9.1.21) from Ref.~\cite{ref:abram} ($J_l$ being the Bessel function of order $l$). Thanks to the above 
we obtain our final expression for the impedance force as
\begin{multline}
F_{y}^{coh}(z;t)= e^{j \Omega t} \frac{j N e^{2}\omega_s \omega_0}{2\pi v \eta Q_{y0}} 
 \sum_{l=-\infty}^{+\infty} j^l \sum_{k=-\infty}^{+\infty} Z_{y}\left(\Omega+k 
\omega_{0}\right) e^{-j \frac{\left(\Omega+k\omega_0\right) z}{v}} \\
\times \int_0^{+\infty} \tilde{r} d \tilde{r} R_{l}(\tilde{r}) J_l\left[\left(\Omega+k\omega_0 - 
\frac{Q_{y}^{\prime}\omega_0}{\eta} \right) \frac{\tilde{r}}{v}\right] . \label{eq:force}
\end{multline}

\subsection{The final equation}

We can finally put into~\Eref{eq:linear_Vlasov_Rl} the coherent force expressed 
in~\Eref{eq:force} to get
\begin{multline}
 \sum_{l'=-\infty}^{+\infty} R_{l'}(r) e^{-j l' \phi} \left(\Omega-Q_{y 0} \omega_{0}-l' 
\omega_{s}\right) = e^{\frac{j Q_{y}^{\prime} r\cos\phi}{\eta R}} g_{0}(r) 
\frac{j N e^{2}\omega_s \omega_0}{4\pi \eta 
Q_{y0} m_{0} \gamma v^2} 
 \sum_{l'=-\infty}^{+\infty} j^{l'} \\
 \times \sum_{k=-\infty}^{+\infty} Z_{y}\left(\Omega+k 
\omega_{0}\right) e^{-j \frac{\left(\Omega+k\omega_0\right) r\cos\phi}{v}} \\
\times \int_0^{+\infty} \tilde{r} d \tilde{r} R_{l'}(\tilde{r}) 
J_{l'}\left[\left(\Omega+k\omega_0 - 
\frac{Q_{y}^{\prime}\omega_0}{\eta} \right) \frac{\tilde{r}}{v}\right] .
\end{multline}
We can get rid of the remaining $\phi$ dependence by trying to expand the right-hand side 
as a Fourier series of $\phi$ and then proceed to term-by-term identification. Both steps 
can be done in one go by integrating both sides of the equation with $\frac{1}{2\pi} \int_0^{2\pi} 
d\phi e^{j l \phi}$. Using~\Eref{eq:Bessel} in the form 
\begin{align}
\frac{1}{2\pi} \int_0^{2\pi} d\phi e^{j l \phi} e^{\frac{j Q_{y}^{\prime} r\cos\phi}{\eta 
R}} e^{-j \left(\Omega+k\omega_0\right) \frac{r\cos\phi}{v}} &=  \frac{1}{2\pi} 
\int_0^{2\pi} d\phi e^{j l \phi} e^{j \left(\frac{j Q_{y}^{\prime} \omega_0}{\eta} - 
\Omega-k\omega_0\right) \frac{r\cos\phi}{v}}  \nonumber \\
&= j^{-l} J_{-l} \left[- \left(\Omega+k\omega_0 - \frac{Q_{y}^{\prime} \omega_0}{\eta} 
\right) \frac{r}{v}\right] \nonumber \\
&= j^{-l} J_{l} \left[ \left(\Omega+k\omega_0 - \frac{Q_{y}^{\prime} \omega_0}{\eta} 
\right) \frac{r}{v}\right],
\end{align}
we obtain
\begin{multline}
R_{l}(r) \left(\Omega-Q_{y 0} \omega_{0}-l 
\omega_{s}\right) = \frac{j N e^{2}\omega_s \omega_0}{4\pi \eta 
Q_{y0} m_{0} \gamma v^2} g_{0}(r)  \sum_{l'=-\infty}^{+\infty} j^{l'-l}
\sum_{k=-\infty}^{+\infty} Z_{y}\left(\Omega+k \omega_{0}\right)  \\
 \times J_{l} \left[ \left(\Omega+k\omega_0 - \frac{Q_{y}^{\prime} \omega_0}{\eta} 
\right) \frac{r}{v}\right]  \int_0^{+\infty} \tilde{r} d \tilde{r} R_{l'}(\tilde{r}) 
J_{l'}\left[\left(\Omega+k\omega_0 - 
\frac{Q_{y}^{\prime}\omega_0}{\eta} \right) \frac{\tilde{r}}{v}\right] . 
\label{eq:Sacherer}
\end{multline}
This is \emph{Sacherer's integral equation}\footnote{In his seminal 
paper~\cite{ref:Sacherer1972}, Sacherer did not write the equation in this form: 
the right-hand side had the form of an integral with a kernel function, and he was using 
the wake function rather than the impedance (see also Eq.~(6.196) of 
Ref.~\cite{ref:Chao}).}.
As we have assumed that $\Omega$ is very close to $Q_{y0} \omega_0 $ and the
impedance is typically a very smooth and slowly varying function, its variation 
between these two frequencies is negligible. The same goes also for the Bessel functions, 
and we can therefore replace all occurrences of $\Omega$ in the right-hand side by 
$Q_{y0} \omega_0$: we get, using again $\omega_0=\frac{v}{R}$
\begin{multline}
R_{l}(r) \left(\Omega-Q_{y 0} \omega_{0}-l 
\omega_{s}\right) = \frac{j N e^{2}\omega_s \omega_0}{4\pi \eta 
Q_{y0} m_{0} \gamma v^2} g_{0}(r)  \sum_{l'=-\infty}^{+\infty} j^{l'-l}
\sum_{k=-\infty}^{+\infty} Z_{y}\left(Q_{y0} \omega_0 + k \omega_{0}\right)  
\\
 \times J_{l} \left[ \left(Q_{y0} + k - \frac{Q_{y}^{\prime}}{\eta} \right) 
\frac{r}{R}\right]  \int_0^{+\infty} \tilde{r} d \tilde{r} R_{l'}(\tilde{r}) 
J_{l'}\left[\left(Q_{y0} + k - \frac{Q_{y}^{\prime}}{\eta} \right) 
\frac{\tilde{r}}{R}\right] . 
\label{eq:Sacherer_noOmega}
\end{multline}
This is essentially Eq.~(6.179) of Ref.~\cite{ref:Chao}. Note that the multiplicative 
coefficient $\frac{j N e^{2}\omega_s \omega_0}{4\pi \eta Q_{y0} m_{0} \gamma v^2}$ on the 
right-hand side, can adopt different forms depending on the normalization conditions 
chosen for the unperturbed distribution $f_0$, which is arbitrary (only the normalization 
of $\psi_0 = f_0 g_0$ is fixed): in Ref.~\cite{ref:Chao} the integral of 
$f_0$ is normalized to $\frac{1}{2\pi}$ instead of $\frac{N}{2\pi}$ here, hence an 
additional factor $N$ here (note also the replacement of all occurrences of $-i$ by $j$
and of $c$ by $v$). In Ref.~\cite[Eq.~(11)]{ref:Mounet_Benevento} $f_0$ was instead 
normalized  to $\frac{N \eta v}{2\pi\omega_s}$, hence a difference 
by a factor $\frac{\eta v}{\omega_s}$ with respect to the above formula\footnote{Note 
also that in the impedance term of Eq.~(11) in Ref.~\cite{ref:Mounet_Benevento}, the 
signs of the expressions between brackets in the Bessel functions have been incorrectly 
inverted: one should read $(\omega_p-\omega_\xi)\tau$ instead of 
$(\omega_\xi-\omega_p)\tau$.}.

\subsection{Solving the final equation}

There are various ways to solve Sacherer's integral equation. Maybe one of the simplest 
is to consider the unperturbed longitudinal distribution to be a Dirac delta function of 
$r$, peaked at a given $r_0$. This is called the \textit{air-bag} model, and its main 
asset is to provide readily the shape of the functions $R_l(r)$ as delta 
functions~\cite[chap.~6]{ref:Chao}. A more sophisticated version of the same idea is to 
discretize $g_0(r)$ as a superposition of such air-bags, each delta function being 
centred at a different $r$ --- this is what is done in the NHTVS~\cite{ref:NHTVS}.

Another method to solve the equation, proposed by Laclare~\cite{ref:Laclare} and used 
also in GALACTIC~\cite{ref:EMetral_GALACTIC} is to integrate over $r$ each side by 
multiplying it with
\[ \int_{0}^{+\infty} r d r J_{l}\left[\left(Q_{y 0}+n-\frac{Q_{y}^{\prime}}{\eta}\right) 
\frac{r}{R}\right],\]
then compute analytically (or numerically, depending on the form of $g_0$)
\[ \int_{0}^{+\infty} r d r g_0(r) J_{l}\left[\left(Q_{y 
0}+n-\frac{Q_{y}^{\prime}}{\eta}\right) \frac{r}{R}\right] J_{l}\left[\left(Q_{y 
0}+k-\frac{Q_{y}^{\prime}}{\eta}\right) \frac{r}{R}\right] ,\]
and finally solve the resulting eigenvalue problem for $\sigma_{ln}$ defined as
\begin{equation}
 \sigma_{ln} \equiv j^l \int_{0}^{+\infty} r d r J_{l}\left[\left(Q_{y 
0}+n-\frac{Q_{y}^{\prime}}{\eta}\right) \frac{r}{R}\right] R_{l}(r) .
\end{equation}
One third class of method is to expand the unknown functions $R_l(r)$ over a basis of 
orthogonal polynomials, e.g. Jacobi polynomials as in Besnier's 
method~\cite{ref:Besnier1984}, or Laguerre polynomials as done in 
MOSES~\cite{ref:MOSES1,ref:MOSES2}, Karliner and Popov's 
approach~\cite{ref:KarlinerPopov}, and in DELPHI~\cite{ref:Mounet_Benevento}:
\begin{equation}
R_{l}(r)=\left(\frac{r}{A}\right)^{|l|} e^{-\kappa r^{2}} \sum_{n=0}^{+\infty} 
c_{l}^{n}L_{n}^{|l|}\left(\kappa r^{2}\right)
\end{equation}
where $L_n^l$ are the associated (or generalized) Laguerre polynomials, $c_l^n$ the 
coefficients of the expansion, and $\kappa$ and A two arbitrary constants that can be 
optimized for each kind of configuration. Integrals involving Bessel functions and 
Laguerre polynomials can be computed analytically, which simplifies greatly the final 
problem. In principle, any family of orthogonal polynomials is appropriate to expand
the radial functions $R_l(r)$; they simply differ in rapidity of convergence of the expansion 
(which can depend on the initial distribution) and whether or not
they allow to compute analytically the involved integrals (which can 
depend on the specific form of the impedance).

Any of the aforementioned method results in an eigenvalue problem of the form
\begin{equation}
\left(\Omega-Q_{y 0} \omega_{0}\right) \alpha_{l 
n}=\sum_{l^{\prime}=-\infty}^{+\infty} \sum_{n^{\prime}=0}^{+\infty}\left(\delta_{l 
l^{\prime}} \delta_{n n^{\prime}} l \omega_{s}+\mathcal{M}_{l n, l^{\prime} 
n^{\prime}}\right) 
\alpha_{l^{\prime} n^{\prime}}
\end{equation}
where $\delta_{k k'}$ is the Kronecker delta (equal to $1$ if $k=k'$, $0$ otherwise), 
the $\alpha_{ln}$ represent the eigenvector looked for (e.g. Laclare's
$\sigma_{ln}$, or the coefficients of the Laguerre expansion $c_l^n$, etc.) and 
$\mathcal{M}$ is a matrix that can be computed 
(semi-)analytically. The eigenvectors represent the mode, expressed according to a 
certain basis that depends on the method chosen to solve the problem. On the other hand, 
the eigenvalue $\Omega-Q_{y 0} \omega_{0}$ always represents the angular 
frequency shift of the mode, and is in practice the main quantity of interest, as the 
sign and magnitude of its imaginary part indicate respectively the unstable character 
(negative sign for an unstable mode) and the growth (or damping) rate of the mode.

So the final equation can be solved simply by diagonalizing $\mathcal{M}$ --- actually a 
partial diagonalization is enough as one typically needs only the most unstable modes, so 
those with the lowest (negative) imaginary part. Note, still, that the matrix is in 
principle infinite and has to be truncated before being diagonalized. The truncation sets 
the number of possible modes and can have an impact even on the most unstable mode if the 
matrix chosen is too small. Therefore one should always check convergence with respect to 
the size of the matrix. 

\section{Conclusion}
\label{sec:conclusion}

In these proceedings we have first described the concepts and basic theory behind Vlasov 
equation, in the context of collective effects in particle accelerators and synchrotrons. 
Then we have set in a general, Hamiltonian framework, 
the application of perturbation theory to Vlasov equation, obtaining its linearized 
version in a compact way, as well as higher order extensions. To that aim, 
Hamiltonians, canonical transformations and the symplecticity condition, as well 
as Poisson brackets, were shortly introduced.

Finally, we have sketched a general method to build a Vlasov solver in practice, applying 
it on the example of transverse instabilities arising from beam-coupling impedance in a 
synchrotron. Most of the emphasis was put on the analytical part of the work, since this 
is typically what one has to do to simplify an \textit{a priori} overly complex partial 
differential equation of seven variables. A number of steps were identified in the 
method, which in that specific case ended-up in Sacherer's integral equation.  We reviewed 
the classical assumptions done to get the well-known equation, which can be summarized in 
the following way:
\begin{itemize}
\item absence of collisions, and of particle diffusion or damping (from e.g. synchrotron radiation),
\item absence of coupling, and of sources of tunespread other than chromaticity,
\item smooth approximation for the single-particle dynamics,
\item the impedance is lumped in a single location, is purely dipolar, and is small 
compared to focusing forces, such that the perturbative treatment is valid,
\item the longitudinal motion is not affected by the transverse impedance nor by chromaticity,
which excludes from the treatment configurations close to transition, low-order 
synchro-betatron resonances, and beams of large transverse size,
\item we can describe the time evolution of the instability with a single exponential that
supersedes all other modes, and its complex coherent frequency of oscillation is 
close to the unperturbed tune,
\item the dependency in the longitudinal and transverse actions of the perturbation to the
distribution function $\Delta \psi$, remain uncoupled.
\end{itemize}

We also described succinctly a few ways to solve this equation; we refer the reader to 
the quoted references for more details.

Although we cannot claim that the approach adopted here is general enough to 
perform equally well in any other context, a few core ideas were highlighted, in 
particular the importance of choosing the right 
canonical coordinates to simplify as much as possible the unperturbed Hamiltonian from 
the beginning, before the linearized Vlasov equation is expanded. Then, the interaction 
from the collective effect under study should 
be specified fully as late as possible, after the perturbation has been already 
simplified. To that aim, decomposing the perturbation of the distribution thanks to 
Fourier series, turned out to be very efficient to reduce the number of variables, as we 
ended-up with a one-dimensional problem that one can solve with a diagonalization.

\section*{Acknowledgements}

I am very grateful to Kevin Li for enlightening discussions on Hamiltonians, Poisson 
brackets, linear operators, and for his elegant derivation of the linearized Vlasov 
equation. I would like to thank as well Sergey Arsenyev, Xavier Buffat and Elias M{\'e}tral 
for proofreading the manuscript and providing numerous helpful comments and suggestions. 
I also thank Sergey Antipov, Giovanni Iadarola, Adrian Oeftiger and 
Beno{\^i}t Salvant for very useful suggestions during the preparation of the lectures. 
My special thanks go to Elias M{\'e}tral and Giovanni Rumolo for motivating the writing of
these proceedings, for their availability for in-depth discussions on the subject, and 
for pointing me to several references.



\providecommand{\href}[2]{#2}\begingroup\raggedright


\begin{thebibliography}{10}

\bibitem{ref:EMetral_summary_LHC_instab_review_runI}
E.~M{\'e}tral {\em et~al.}, Summary of the 2-day internal review of {LHC}
  performance limitations (linked to transverse collective effects) during run
  {I} ({CERN}, 25-26/09/2013), CERN-ACC-NOTE-2014-0006 (2014),
\newblock
  {\url{https://cds.cern.ch/record/1645854/files/CERN-ACC-NOTE-2014-0006.pdf}}.

\bibitem{ref:EMetral_summary_LHC_instab_review_runII}
E.~M{\'e}tral {\em et~al.}, Summary of the half-day internal review of {LHC}
  performance limitations (linked to transverse collective effects) during run
  {II} ({CERN}, 29/11/2016), CERN-ACC-NOTE-2017-0005 (2017),
\newblock
  {\url{https://cds.cern.ch/record/2242249/files/CERN-ACC-NOTE-2017-0005.pdf}}.

\bibitem{ref:IEEE}
E.~M{\'e}tral {\em et~al.},
\newblock {\em IEEE Trans. Nucl. Sci.} {\bfseries 63}, 2  (2016) 1001,
\newblock {\url{https://doi.org/10.1109/TNS.2015.2513752}}.

\bibitem{ref:Salvant_Benevento}
B.~Salvant,
\newblock Impedance and instabilities in hadron machines,
\newblock in Proc. ICFA Mini-Workshop on Impedances and Beam Instabilities in
  Particle Accelerators, Benevento, Italy, 18--22 September 2017, Eds.
  V.~Brancolini, G.~Rumolo, M.~R. Masullo, and S.~Petracca, CERN-2018-003-CP
  (CERN, Geneva, 2018),  Vol.~1, p.~99,
\newblock {\url{https://doi.org/10.23732/CYRCP-2018-001.99}}.

\bibitem{ref:Laslett}
L.~J. Laslett, V.~K. Neil, and A.~M. Sessler,
\newblock {\em Rev. Sci. Instrum.} {\bfseries 36}, 4  (1965) 436,
\newblock {\url{https://doi.org/10.1063/1.1719595}}.

\bibitem{ref:Sacherer1972}
F.~J. Sacherer, Methods for computing bunched-beam instabilities,
  CERN/SI-BR/72-5 (1972),
\newblock {\url{https://cds.cern.ch/record/322545/files/CM-P00063598.pdf}}.

\bibitem{ref:Kli}
K.~Li,
\newblock Numerical methods {I} \& {II},
\newblock in Proc. of the CAS-CERN Accelerator School on Intensity Limitations
  in Particle Beams, Geneva, Switzerland, 2--11 November 2015, Ed. W.~Herr,
  CERN-2017-006-SP (CERN, Geneva, 2017),  Vol.~3, p. 247,
\newblock {\url{https://doi.org/10.23730/CYRSP-2017-003.247}}.

\bibitem{ref:Migliorati}
M.~Migliorati and D.~Quartullo,
\newblock Impedance-induced beam instabilities and damping mechanisms in
  circular machines - longitudinal - simulations,
\newblock in ICFA Beam Dynamics Newsletter, edited by E.~M{\'e}tral,  No. ~69
  (2016), p. 132,
\newblock {\url{https://cds.cern.ch/record/2244873/files/ICFA69_132-141.pdf}}.

\bibitem{ref:KLi_MSchenk}
K.~Li and M.~Schenk,
\newblock Impedance-induced beam instabilities and damping mechanisms in
  circular machines - transverse - simulations,
\newblock in ICFA Beam Dynamics Newsletter, edited by E.~M{\'e}tral,  No. ~69
  (2016), p. 177,
\newblock {\url{https://cds.cern.ch/record/2264412/files/ICFA69_177-185.pdf}}.

\bibitem{ref:Mounet_Benevento}
N.~Mounet,
\newblock Vlasov solvers and macroparticle simulations,
\newblock in Proc. ICFA Mini-Workshop on Impedances and Beam Instabilities in
  Particle Accelerators, Benevento, Italy, 18--22 September 2017, Eds.
  V.~Brancolini, G.~Rumolo, M.~R. Masullo, and S.~Petracca, CERN-2018-003-CP
  (CERN, Geneva, 2018),  Vol.~1, p.~77,
\newblock {\url{https://doi.org/10.23732/CYRCP-2018-001.77}}.

\bibitem{ref:Chen}
F.~F. Chen,
\newblock {\em Introduction to Plasma Physics and Controlled Fusion}, 3rd ed.
  (Springer, 2016),
\newblock {\url{https://doi.org/10.1007/978-3-319-22309-4}}.

\bibitem{ref:Chao}
A.~W. Chao,
\newblock {\em Physics of Collective Beams Instabilities in High Energy
  Accelerators} (John Wiley \& Sons, 1993),
\newblock {\url{http://www.slac.stanford.edu/~achao/wileybook.html}}.

\bibitem{ref:Vlasov_in_Russian}
A.~A. Vlasov,
\newblock {\em Zh. Ehksp. Teor. Fiz.} {\bfseries 8}, 3  (1938) 291, in
  {R}ussian.

\bibitem{ref:Vlasov}
A.~A. Vlasov,
\newblock {\em Russ. Phys. J.} {\bfseries 9}, 1  (1945) 25.

\bibitem{ref:Chandrasekhar}
S.~Chandrasekhar,
\newblock {\em Rev. Mod. Phys.} {\bfseries 15}, 1  (1943) 1,
\newblock {\url{https://doi.org/10.1103/RevModPhys.15.1}}.

\bibitem{ref:WarnockVlasovFokkerPlanck2006}
R.~L. Warnock,
\newblock {\em Nucl. Instrum. Methods Phys. Res. A} {\bfseries 561}, 2  (2006)
  186,
\newblock {\url{https://doi.org/10.1016/j.nima.2006.01.041}}.

\bibitem{ref:Lindberg}
R.~R. Lindberg,
\newblock {\em Phys. Rev. Accel. Beams} {\bfseries 19}, 12  (2016) 124402,
\newblock {\url{https://doi.org/10.1103/PhysRevAccelBeams.19.124402}}.

\bibitem{ref:Chao1984}
A.~W. Chao and R.~D. Ruth,
\newblock {\em Part. Accel.} {\bfseries 16} (1985) 201,
\newblock {\url{https://cds.cern.ch/record/156021/files/p201.pdf}}.

\bibitem{ref:Alexahin}
Y.~Alexahin,
\newblock {\em Nucl. Instrum. Methods Phys. Res. A} {\bfseries 480}, 2  (2002)
  253,
\newblock {\url{https://doi.org/10.1016/S0168-9002(01)01219-0}}.

\bibitem{ref:Perevedentsev}
E.~A. Perevedentsev,
\newblock Head-tail instability caused by electron cloud,
\newblock in Proc. Mini Workshop on Electron Cloud Simulations for Proton and
  Positron Beams, CERN, Geneva, Switzerland, 15--18 April 2002, Eds. G.~Rumolo
  and F.~Zimmermann, CERN-2002-001 (CERN, Geneva, 2002), p. 171,
\newblock {\url{https://doi.org/10.5170/CERN-2002-001.171}}.

\bibitem{ref:Davidson}
R.~C. Davidson, H.~Qin, and T.-S.~F. Wang,
\newblock {\em Phys. Lett.~A} {\bfseries 252}, 5  (1999) 213,
\newblock {\url{https://doi.org/10.1016/S0375-9601(99)00002-X}}.

\bibitem{ref:Ryne}
R.~D. Ryne and A.~J. Dragt,
\newblock Lie algebraic treatment of space charge,
\newblock in Proc. 12th IEEE Particle Accelerator Conference, Washington, DC,
  USA, 16--19 March 1987, Eds. E.~R. Lindstrom and L.~S. Taylor (IEEE, New
  York, 1987), p. 1063,
\newblock
  {\url{http://accelconf.web.cern.ch/accelconf/p87/PDF/PAC1987_1063.PDF}}.

\bibitem{ref:Blaskiewicz}
M.~Blaskiewicz,
\newblock {\em Phys. Rev. Spec. Top. Accel. Beams} {\bfseries 1}, 4  (1998)
  044201,
\newblock {\url{https://doi.org/10.1103/PhysRevSTAB.1.044201}}.

\bibitem{ref:Burov}
A.~Burov,
\newblock {\em Phys. Rev. Spec. Top. Accel. Beams} {\bfseries 12}, 4  (2009)
  044202,
\newblock {\url{https://doi.org/10.1103/PhysRevSTAB.12.044202}}.

\bibitem{ref:Balbekov}
V.~Balbekov,
\newblock {\em Phys. Rev. Spec. Top. Accel. Beams} {\bfseries 12}, 12  (2009)
  124402,
\newblock {\url{https://doi.org/10.1103/PhysRevSTAB.12.124402}}.

\bibitem{ref:Shobuda}
Y.~Shobuda {\em et~al.},
\newblock {\em Prog. Theor. Exp. Phys.} {\bfseries 2017}, 1  (2017) 013G01,
\newblock {\url{https://doi.org/10.1093/ptep/ptw169}}.

\bibitem{ref:Venturini}
M.~Venturini, R.~Warnock, and A.~Zholents,
\newblock {\em Phys. Rev. Spec. Top. Accel. Beams} {\bfseries 10}, 5  (2007)
  054403,
\newblock {\url{https://doi.org/10.1103/PhysRevSTAB.10.054403}}.

\bibitem{ref:Lindberg_longitudinal}
R.~R. Lindberg,
\newblock {\em Phys. Rev. Accel. Beams} {\bfseries 21}, 12  (2018) 124402,
\newblock {\url{https://doi.org/10.1103/PhysRevAccelBeams.21.124402}}.

\bibitem{ref:Sacherer1974}
F.~J. Sacherer,
\newblock Transverse bunched beam instabilitites - {T}heory,
\newblock in Proc. 9th International Conference on High-energy Accelerators,
  SLAC, Stanford, California, USA, 2--7 May 1974, CERN-MPS-INT-BR-74-8,
  SLAC-REPRINT-1974-004, CONF-74-0522 (A.E.C., Washington, DC, 1975), p. 347,
\newblock {\url{https://cds.cern.ch/record/322645/files/HEACC74_368-372.pdf}}.

\bibitem{ref:Zotter_EMajorana}
B.~Zotter,
\newblock Transverse instabilities of relativistic particle beams in
  accelerators and storage rings -- {P}art {I}: Coasting beams,
\newblock in Proc. of the 1st International School of Particle Accelerators
  ``Ettore Majorana'', Erice, Italy, 10--22 November 1976, Ed. M.~H. Blewett,
  CERN-77-13 (CERN, Geneva, 1977), p. 175,
\newblock {\url{https://doi.org/10.5170/CERN-1977-013.175}}.

\bibitem{ref:Sacherer_EMajorana}
F.~J. Sacherer,
\newblock Transverse instabilities of relativistic particle beams in
  accelerators and storage rings -- {P}art {II}: Bunched beams,
\newblock in Proc. of the 1st International School of Particle Accelerators
  ``Ettore Majorana'', Erice, Italy, 10--22 November 1976, Ed. M.~H. Blewett,
  CERN-77-13 (CERN, Geneva, 1977), p. 198,
\newblock {\url{https://doi.org/10.5170/CERN-1977-013.198}}.

\bibitem{ref:Besnier1979}
G.~Besnier,
\newblock {\em Nucl. Instrum. Methods} {\bfseries 164}, 2  (1979) 235,
\newblock {\url{https://doi.org/10.1016/0029-554X(79)90241-6}}.

\bibitem{ref:Besnier1984}
G.~Besnier, D.~Brandt, and B.~Zotter,
\newblock {\em Part. Accel.} {\bfseries 17} (1985) 51,
\newblock {\url{https://cds.cern.ch/record/154782/files/p51.pdf}}.

\bibitem{ref:Chin_LandauDamping}
Y.-H. Chin, Hamiltonian formulation for transverse bunched beam instabilities
  in the presence of betatron tunespread, CERN-SPS-85-9-DI-MST (1985),
\newblock {\url{https://cds.cern.ch/record/160217/files/198507010.pdf}}.

\bibitem{ref:Laclare}
J.~L. Laclare,
\newblock Bunched beam coherent instabilities,
\newblock in Proc. of the CAS-CERN Accelerator School: Accelerator Physics,
  Oxford, UK, 16--27 September 1985, Ed. S.~Turner, CERN-1987-003-V-1 (CERN,
  Geneva, 1987), p. 264,
\newblock {\url{https://doi.org/10.5170/CERN-1987-003-V-1.264}}.

\bibitem{ref:JSBerg}
J.~Scott~Berg,
\newblock Ph.D. thesis, Stanford University, 1996,
\newblock {\url{http://www.slac.stanford.edu/cgi-wrap/getdoc/slac-r-478.pdf}}.

\bibitem{ref:KarlinerPopov}
M.~Karliner and K.~Popov,
\newblock {\em Nucl. Instrum. Methods Phys. Res. A} {\bfseries 537}, 3  (2005)
  481,
\newblock {\url{https://doi.org/10.1016/j.nima.2004.08.068}}.

\bibitem{ref:Schenk}
M.~Schenk, X.~Buffat, K.~Li, and A.~Maillard,
\newblock {\em Phys. Rev. Spec. Top. Accel. Beams} {\bfseries 21}, 8  (2018)
  084402,
\newblock {\url{https://doi.org/10.1103/PhysRevAccelBeams.21.084402}}.

\bibitem{ref:EMetral_GALACTIC_GALACLIC}
E.~M{\'e}tral,
\newblock {GALACTIC and GALACLIC: two Vlasov solvers for the transverse and
  longitudinal planes},
\newblock in Proc. 10th International Particle Accelerator Conference (IPAC
  2019), Melbourne, Australia, 19--24 May 2019, Eds. M.~Boland {\em et~al.}
  (JACoW Publishing, 2019), MOPGW087,
\newblock {\url{https://ipac2019.vrws.de/papers/mopgw087.pdf}}.

\bibitem{ref:MOSES1}
Y.-H. Chin, Transverse mode coupling instabilities in the {SPS},
  CERN-SPS-85-2-DI-MST (1985),
\newblock {\url{https://cds.cern.ch/record/157995/files/198503163.pdf}}.

\bibitem{ref:MOSES2}
Y.-H. Chin, User's guide for new {MOSES} version 2.0: Mode-coupling single
  bunch instability in an electron storage ring, CERN-LEP-TH-88-05 (1988),
\newblock {\url{https://cds.cern.ch/record/187253/files/198805308.pdf}}.

\bibitem{ref:NHTVS}
A.~V. Burov,
\newblock {\em Phys. Rev. Spec. Top. Accel. Beams} {\bfseries 17}, 021007
  (2014) 021007,
\newblock {\url{https://doi.org/10.1103/PhysRevSTAB.17.021007}}.

\bibitem{ref:EMetral_GALACTIC}
E.~M{\'e}tral {\em et~al.},
\newblock Destabilising effect of the {LHC} transverse damper,
\newblock in Proc. 9th International Particle Accelerator Conference (IPAC
  2018), Vancouver, BC, Canada, 29 April -- 4 May 2018, Eds. S.~Koscielniak,
  T.~Satogata, V.~Schaa, and J.~Thomson, [{\em J. Phys. Conf. Ser.} {\bfseries
  1067}, 011001  (2018) THPAF048],
\newblock {\url{https://doi.org/10.18429/JACoW-IPAC2018-THPAF048}}.

\bibitem{ref:KLi_Vlasov}
K.~Li,
\newblock Perturbation formalism,
\newblock Lecture at the US Particle Accelerator School, Hampton, Virginia,
  USA, 19--23 January 2015,
\newblock
  {\url{http://kli.web.cern.ch/kli/USPAS_Lectures_Collective_Effects/Lectures/USPAS_05a_Vlasov_equation.pdf}}.

\bibitem{ref:Goldstein}
H.~Goldstein, C.~P. Poole~Jr., and J.~L. Safko,
\newblock {\em Classical Mechanics}, 3rd ed. (Addison Wesley, San Francisco,
  California, 2002).

\bibitem{ref:Bell}
J.~S. Bell,
\newblock Hamiltonian mechanics,
\newblock in Proc. of the CAS-CERN Accelerator School: Accelerator Physics,
  Oxford, UK, 16--27 September 1985, Ed. S.~Turner, CERN-1987-003-V-1 (CERN,
  Geneva, 1987), p.~5,
\newblock {\url{https://doi.org/10.5170/CERN-1987-003-V-1.5}}.

\bibitem{ref:Ruth_Hamiltonian}
R.~D. Ruth,
\newblock Single-particle dynamics in circular accelerators,
\newblock in Proc. of the SLAC Summer School on the Physics of High-energy
  Particle Accelerators, Standford, California, USA, 15--26 July 1985, Eds.
  M.~Month and M.~Dienes, [{\em AIP Conf. Proc.} {\bfseries 153} (1987) 150],
\newblock {\url{https://doi.org/10.1063/1.36365}}.

\bibitem{ref:Herr}
W.~Herr,
\newblock Mathematical and numerical methods for non-linear beam dynamics,
\newblock in Proc. of the CAS-CERN Accelerator School: Advanced Accelerator
  Physics, Trondheim, Norway, 18--29 August 2013, Ed. W.~Herr, CERN-2014-009
  (CERN, Geneva, 2014), p. 157,
\newblock {\url{https://doi.org/10.5170/CERN-2014-009.157}}.

\bibitem{ref:Zotter}
A.~Chao, S.~Heifets, and B.~Zotter,
\newblock {\em Phys. Rev. Spec. Top. Accel. Beams} {\bfseries 5}, 11  (2002)
  111001,
\newblock {\url{https://doi.org/10.1103/PhysRevSTAB.5.111001}}.

\bibitem{ref:Bane_wake}
K.~L.~F. Bane, P.~B. Wilson, and T.~Weiland,
\newblock Wake fields and wake field acceleration,
\newblock in Proc. of the 3rd Summer School on High-Energy Particle
  Accelerators, Upton, New York, USA, 6--16 July 1983, Eds. M.~Month, P.~F.
  Dahl, and M.~Dienes, [{\em AIP Conf. Proc.} {\bfseries 127} (1985) 875],
\newblock {\url{https://doi.org/10.1063/1.35182}}.

\bibitem{ref:abram}
M.~Abramowitz and I.~A. Stegun, editors,
\newblock {\em Handbook of Mathematical Functions With Formulas, Graphs, and
  Mathematical Tables} (US Dep. of Commerce, National Bureau of Standards,
  1972), Tenth Printing, with corrections,
\newblock
  {\url{http://www.iopb.res.in/~somen/abramowitz\_and\_stegun/toc.htm}}.

\end{thebibliography}

\appendix
\section{Canonical transformation to action -- angles coordinates}
\label{app:symplectic}

To show that the transformation of coordinates \[ (y,y',z,\delta) \rightarrow 
(J_y,\theta_y,J_z,\phi),\] used in Section~\ref{sec:coord} is canonical, we can show 
separately that both the transformations $(y,y') \; \rightarrow \; (J_y,\theta_y)$ 
and $(z,\delta) \; \rightarrow \; (J_z,\phi)$ are canonical, since these two 
transformations of coordinates are uncoupled.

For the transformation $(y,y') \; \rightarrow \; (J_y,\theta_y)$, which is expressed as
\begin{align}
J_{y} &=\frac{1}{2}\left( \frac{Q_{y0}}{R} y^{2} + \frac{R}{Q_{y0}} y^{\prime 2} \right), 
\qquad  \theta_{y} =\operatorname{atan}\left(\frac{R y^{\prime}}{Q_{y0} y}\right), 
\end{align}
we have for the Jacobian
\begin{equation}
\mathcal{J}=\begin{pmatrix}{\frac{\partial J_{y}}{\partial y}} & 
{\frac{\partial J_{y}}{\partial y^{\prime}}} \\ \\ {\frac{\partial \theta_{y}}{\partial 
y}} & {\frac{\partial \theta_{y}}{\partial y^{\prime}}}\end{pmatrix}=
\begin{pmatrix}{\frac{y Q_{y0}}{R}} & {\frac{y^{\prime} R}{Q_{y0}}} \\ \\ {- 
\frac{y'}{2J_y}} & {\frac{y}{2J_y}} \end{pmatrix},
\end{equation}
such that
\begin{align}
\mathcal{J}^T \cdot \begin{pmatrix}{0} & {1} \\ {-1} & {0}\end{pmatrix} \cdot \mathcal{J} 
&= \begin{pmatrix}{\frac{y Q_{y0}}{R}} & {-\frac{y'}{2J_y}} \\ \\ 
{\frac{y^{\prime} R}{Q_{y0}}} & {\frac{y}{2J_y}}\end{pmatrix} \cdot \begin{pmatrix}{0} & {1} \\ {-1} & {0}\end{pmatrix} \cdot 
\begin{pmatrix}{\frac{y Q_{y0}}{R}} & {\frac{y^{\prime} R}{Q_{y0}}} \\ \\ 
{-\frac{y'}{2J_y}} & {\frac{y}{2J_y}}\end{pmatrix} \nonumber \\ \nonumber \\
&=\begin{pmatrix}{\frac{y'}{2J_y}} & {\frac{y 
Q_{y0}}{R}} \\ \\ {-\frac{y}{2J_y}} & {\frac{y^{\prime} 
R}{Q_{y0}}}\end{pmatrix} \cdot \begin{pmatrix}{\frac{y Q_{y0}}{R}} & 
{\frac{y^{\prime} R}{Q_{y0}}} \\ \\ {-\frac{y'}{2J_y}} & 
{\frac{y}{2J_y}}\end{pmatrix} \nonumber \\ 
\nonumber \\
&=\begin{pmatrix}{0} & {1} \\ {-1} & {0}\end{pmatrix},
\end{align}
which is the symplecticity condition, hence the transformation $(y,y') \; \rightarrow \; 
(J_y,\theta_y)$ is canonical.

Then, for the transformation $(z,\delta) \; \rightarrow \; (J_z,\phi)$ we have in turn
\begin{align}
 J_{z} &=\frac{1}{2}\left(\frac{\omega_{s}}{v \eta} z^{2}+\frac{v \eta}{\omega_{s}}
\delta^{2}\right), \qquad  \phi =\operatorname{atan}\left(\frac{v \eta 
\delta}{\omega_{s} z}\right), 
\end{align}
such that the Jacobian is
\begin{equation}
\mathcal{J}=\begin{pmatrix}{\frac{\partial J_{z}}{\partial z}} & 
{\frac{\partial J_{z}}{\partial \delta}} \\ \\ {\frac{\partial \phi}{\partial z}} & 
{\frac{\partial \phi}{\partial \delta}}\end{pmatrix}= 
\begin{pmatrix}{\frac{\omega_{s}}{v \eta} z} & {\frac{v \eta}{\omega_{s}} \delta} \\ \\ 
{-\frac{\delta}{2 J_{z}}} & {\frac{z}{2 J_{z}}}\end{pmatrix},
\end{equation}
and we get
\begin{align}
\mathcal{J}^T \cdot \begin{pmatrix}{0} & {1} \\ {-1} & {0}\end{pmatrix} \cdot \mathcal{J} 
&= \begin{pmatrix}{\frac{\omega_{s}}{v \eta} z} & {-\frac{\delta}{2 J_{z}}} \\ \\ 
{\frac{v \eta}{\omega_{s}} \delta} & {\frac{z}{2 J_{z}}}\end{pmatrix} \cdot 
\begin{pmatrix}{0} & {1} \\ {-1} & {0}\end{pmatrix} \cdot 
\begin{pmatrix}{\frac{\omega_{s}}{v \eta} z} & {\frac{v \eta}{\omega_{s}} \delta} \\ \\ 
{-\frac{\delta}{2 J_{z}}} & {\frac{z}{2 J_{z}}}\end{pmatrix} \nonumber \\ \nonumber \\
 &=\begin{pmatrix}{\frac{\delta}{2 J_z}} & {\frac{\omega_{s}}{v \eta} z} \\ \\ 
{-\frac{z}{2 J_{z}}} & {\frac{v \eta}{\omega_{s}} \delta}\end{pmatrix} \cdot 
\begin{pmatrix}{\frac{\omega_{s}}{v \eta} z} & {\frac{v \eta}{\omega_{s}} 
\delta} \\ \\ {-\frac{\delta}{2 J_{z}}} & {\frac{z}{2 J_{z}}}\end{pmatrix} \nonumber \\ 
\nonumber \\
 &= \begin{pmatrix}{0} & {1} \\ {-1} & {0}\end{pmatrix}.
\end{align}
Therefore, the full transformation of coordinates $(y,y',z,\delta) \rightarrow 
(J_y,\theta_y,J_z,\phi)$ is canonical.

\end{document}